\documentclass[prd,superscriptaddress,amsfonts,amssymb,amsmath,showpacs]{revtex4-2}
\usepackage{bm}
\usepackage{amsfonts}
\usepackage{latexsym}
\usepackage[latin1]{inputenc}
\usepackage{graphicx}
\usepackage{amsmath}
\usepackage{palatino}
\usepackage{mathpazo}
\usepackage[british]{babel}
\usepackage{hhline}
\usepackage{multirow}
\usepackage{textcomp}
\usepackage{float}
\usepackage{booktabs}
\usepackage{dcolumn}
\usepackage{ragged2e}
\usepackage{hyperref}
\hypersetup{colorlinks,citecolor=blue}
\hypersetup{colorlinks=true,linkcolor=red,filecolor=magenta,    urlcolor=cyan}
\usepackage{amsmath}
\usepackage{xcolor}
\usepackage{orcidlink}
\usepackage{epsfig}
\usepackage{caption}
\usepackage{subcaption}
\usepackage{commath}
\def\jnl@style{\it}
\def\aaref@jnl#1{{\jnl@style#1}}

\def\aaref@jnl#1{{\jnl@style#1}}

\def\aj{\aaref@jnl{AJ}}                   
\def\apj{\aaref@jnl{ApJ}}                 
\def\apjl{\aaref@jnl{ApJ}}                
\def\apjs{\aaref@jnl{ApJS}}               
\def\apss{\aaref@jnl{Ap\&SS}}             
\def\aap{\aaref@jnl{A\&A}}                
\def\aapr{\aaref@jnl{A\&A~Rev.}}          
\def\aaps{\aaref@jnl{A\&AS}}              
\def\mnras{\aaref@jnl{Mon.~Not.~Roy.~Astron.~Soc.}}             
\def\prd{\aaref@jnl{Phys.~Rev.~D}}        
\def\prc{\aaref@jnl{Phys.~Rev.~C}}  
\def\prl{\aaref@jnl{Phys.~Rev.~Lett.}}    
\def\qjras{\aaref@jnl{QJRAS}}             
\def\skytel{\aaref@jnl{S\&T}}             
\def\ssr{\aaref@jnl{Space~Sci.~Rev.}}     
\def\zap{\aaref@jnl{ZAp}}                 
\def\nat{\aaref@jnl{Nature}}              
\def\aplett{\aaref@jnl{Astrophys.~Lett.}} 
\def\apspr{\aaref@jnl{Astrophys.~Space~Phys.~Res.}} 
\def\physrep{\aaref@jnl{Phys.~Rep.}}      
\def\physscr{\aaref@jnl{Phys.~Scr}}       
\def\commat{\aaref@jnl{Comm.~Math.~Phys.}}              
\def\science{\aaref@jnl{Science}}               
\def\cqg{\aaref@jnl{Classical Quant.~Grav.}}            
\def\jpcs{\aaref@jnl{JPCS}}                                     
\def\ijmpd{\aaref@jnl{Int.~J.~Mod.~Phys.~D}}                    
\def\grg{\aaref@jnl{Gen.~Relat.~Gravit.}}               
\def\rpp{\aaref@jnl{Rep.~Prog.~Phys.}}          
\def\npa{\aaref@jnl{Nucl.~Phys.~A}}        
\def\lrr{\aaref@jnl{Living Rev.~Rel.}}                   
\def\jcap{\aaref@jnl{J.~Cosmology Astropart.~Phys.}}    
\def\rmp{\aaref@jnl{Rev.~Mod.~Phys.}}   
\def\epjc{\aaref@jnl{Eur.~Phys.~J.~C}} 
\def\plb{\aaref@jnl{~Phy.~Lett.~B}} 
\def\mpla{\aaref@jnl{Mod.~Phy.~Lett.~A}} 
\def\arxiv{\aaref@jnl{arxiv.org}}

\allowdisplaybreaks[1]

\begin{document}
\color{black}       
\title{\bf Bouncing cosmology in modified gravity with higher-order Gauss-Bonnet curvature term}

\author{Santosh V Lohakare \orcidlink{0000-0001-5934-3428}}
\email{lohakaresv@gmail.com}
\affiliation{Department of Mathematics,
Birla Institute of Technology and Science-Pilani, Hyderabad Campus,
Hyderabad-500078, India.}

\author{Francisco Tello-Ortiz \orcidlink{0000-0002-7104-5746}}
\email{francisco.tello@ua.cl}
\affiliation{Departamento de F\'isica, Facultad de Ciencias B\'asicas, Universidad de Antofagasta, Casilla 170, Antofagasta, Chile.}

\author{S.K. Tripathy \orcidlink{0000-0001-5154-2297}} 
\email{tripathy$\_$sunil@rediffmail.com} 
\affiliation{Department of Physics, Indira Gandhi Institute of Technology, Sarang, Dhenkanal, Odisha-759146, India.}

\author{B. Mishra \orcidlink{0000-0001-5527-3565}}
\email{bivu@hyderabad.bits-pilani.ac.in}
\affiliation{Department of Mathematics,
Birla Institute of Technology and Science-Pilani, Hyderabad Campus,
Hyderabad-500078, India.}

\begin{abstract}
\textbf{Abstract}: In this paper, we have studied the bouncing behaviour of the cosmological models formulated at the background of Hubble function in $F(R, \mathcal{G})$ theory of gravity, where $R$ and $\mathcal{G}$ respectively denotes the Ricci scalar and Gauss-Bonnet invariant. The actions of bouncing cosmology are studied under consideration of different viable models and can resolve the difficulty of singularity in standard Big-Bang cosmology. Both the models are showing bouncing behaviour and satisfying the bouncing cosmological properties. Models based on dynamical, deceleration, and energy conditions indicate the accelerating behavior at the late evolution time. Phantom at the bounce epoch is analogous to a quintessence behavior. Finally, we formulate the perturbed evolution equations and investigate the stability of two bouncing solutions.
\end{abstract}

\maketitle

\section{Introduction} \label{SEC I}
In recent astrophysics and cosmology research, instances of late-time cosmic acceleration supposedly witnessed by cosmological observations have compelled theoretical cosmologists and astrophysicists to think beyond general relativity (GR)~
. Cosmological observations such as  those of high redshift supernovae \cite{Riess98}, supernovae of type Ia \cite{Perlmutter99, Bennett03}, cosmic microwave background radiations (CMBRs) \cite{Spergel03, Spergel07}, baryon acoustic oscillations \cite{Percival10}, and Planck collaboration \cite{Ade14} are sufficiently indicative of  the accelerated expansion of the universe. Furthermore, the cause has been speculated to be the presence of some exotic dark energy (DE). The negative pressure indicates the violation of strong energy conditions as well as a limitation in GR. Accordingly, in the field equations of GR, the modification is thought to be in the geometrical part or matter part. The matter can be modified by replacing the dynamical parameters with the DE parameters that lead to the DE models. The modification in the geometrical part can be performed by including additional terms, known as geometrically extended gravity models. Some recent geometrical extended gravities are $F(R)$ gravity \cite{Nojiri03a,Carroll04,Barragan09}, $F(R,T)$ gravity \cite{Harko11}, $F(\mathcal{T})$ gravity~\cite{Cai11,Abedi18}, $F(Q,T)$ gravity \cite{Xu19}, etc. Another such extension is made by including the Gauss--Bonnet invariant, known as the $F(R,\mathcal{G})$ gravity \cite{Nojiri05}. The next section discusses $F(R,\mathcal{G})$ gravity in detail.

The standard model or Big Bang cosmology has been widely accepted as a cosmological model and has successfully defended many intriguing universe problems. However, regarding the issue of late-time cosmic acceleration, GR has difficulties in resolving some early universe issues such as an initial singularity, flatness, and cosmic horizon. The inflationary scenario could resolve the flatness and horizon issues, but not that of initial singularity. To date, the beginning of the universe before inflation is not known. This is because of the attractor nature of inflation. When the inflation started, the information on the initial singularity was lost as the initial spatial curvature was stretched away by the exponential expansion of inflation. The matter bounce scenario can remove the initial singularity. To achieve this, new physics is required to supply the bounce, which can be obtained by introducing new kinds of matter such as phantom or quintom fields \cite{Cai07}, ghost condensates \cite{Peter02,Lin11}, effective string theory actions~\cite{Fabris03}, Galileons~\cite{Qiu11}, and S-branes \cite{Kounnas12}. In the matter bounce scenario, the matter controls the contraction and can generate early-density perturbations with a nearly scale-invariant and adiabatic spectrum \cite{Novello08,Elizalde15}. Cosmological models before inflationary cosmology produce scale invariance and a nearly adiabatic spectrum of cosmological perturbation that agree with the observations made in \cite{Peebles70,Sunyaev70}. Under the purview of GR, the bouncing behavior cannot be achieved since it leads to the violation of the null energy condition, which is happening because of the change in sign of the astrophysicist rate at the bounce objective \cite{Battefeld15}. It is important to note that the violation of the null energy condition is not inheritable from modified gravity \cite{Novello08} or quantum mechanics \cite{Barrau17}. Several such models are suggested in the literature, such as loop quantum gravity \cite{Bojowald01,Odintsov14}, Ekpyrotic models \cite{Khoury01}, Pre-Big Bang \cite{Gasperini03}, gravity actions with higher-order corrections \cite{Nojiri03}, braneworld scenarios \cite{Saridakis09}, non-relativistic gravity \cite{Brandenberger09,Cai09}, $F(\mathcal{T})$ gravity \cite{Cai11}, and $F(Q,T)$ gravity \cite{Agrawal21}.

In bouncing cosmology, Cai \cite{Cai16} has observationally shown that the matter bounce scenario allowed for a sizeable parameter space when the background energy density was small. Shabani and Ziaie \cite{Shabani18} have studied the classical bounce solution in $F(R,T)$ gravity. In the different framework of $F(R, T)$ gravity, the matter bounce cosmology has been widely studied  \cite{Mishra19, Tripathy21}. The bouncing scenario is designed to avoid the Big Bang singularity in the background of isotropic and anisotropic space--time examined by Agrawal et al.~\cite{Agrawal22p, Agrawal22f}. Amani \cite{Amani16} has reconstructed the $F(R)$ gravity by the redshift parameter and has shown the bouncing behavior at the backdrop of homogeneous and isotropic space--time. Nojiri et al. \cite{Nojiri16} have demonstrated the realization of bouncing cosmology from unimodular $F(R)$ gravity. Ilyas and Rahman \cite{Ilyas21} have rebuilt the model with a redshift parameter to present the bouncing scenario in $F(R)$ gravity. In $F(\mathcal{T})$ gravity, Amoros et al. \cite{Amoros13} studied the bouncing behavior from loop quantum cosmology. In $F(\mathcal{T}, B)$ gravity, Caruana et al. \cite{Caruana20} gave the cosmological bouncing solution with a power law cosmology.  We can infer that the modified theories of gravity successfully resolved the initial singularity issue through bouncing cosmology. In the present study, we are interested in studying the bouncing behavior within a modified $F(R,\mathcal{G})$ gravity.

{There are several bouncing scale factors available in the literature \cite{Odintsov20}. Among these, the scale factor $a(t)=(a_0 t^2+1)^n$ is of special significance. Through this scale factor, we can solve the singularity issue. 
In our work, we have introduced this scale factor and resolved the bounce issue. Apart from this, it has some other advantages such that the slow roll conditions that are assumed to hold; thus, the observational indices have general expressions regarding the slow-roll parameters in the inflationary scenario investigated by Odintsov et al. \cite{Odintsov20}. Using modified teleparallel gravity, Karimzadeh et al. demonstrated the possibility of achieving effective phantom behavior without adopting phantom fields \cite{Karimzadeh19}. The scalar perturbation technique creates the perturbed evolution equations, and their stability has been demonstrated by Duchaniya et al. \cite{Duchaniya22}.}

The plan of this paper is as follows. In Section \ref{SEC II}, we discuss $F(R,\mathcal{G})$ gravity and its field equations are presented along with the dynamical parameters. In Sections \ref{SEC III} and \ref{SEC IV}, the cosmological model is constructed using the bouncing scale factor determined by the physical parameters. Within the bouncing scenario, different dynamical parameters are discovered. The energy conditions of the models were analyzed. A linear, homogeneous, and isotropic perturbation computation was performed in Section \ref{SEC V} to assess the stability of the model. Finally, Section \ref{SEC VI} presents the conclusion of the work.

\section{$F(R, \mathcal{G})$ Gravity and Field Equations} \label{SEC II}
Generalizations of $F(R)$ and $F(\mathcal{G})$ gravity are offered by higher-order gravities~\cite{Barrow88,Capozziello04}. This uses combinations of higher-order curvature invariants constructed from the Ricci tensor $R_{\mu \nu}$ and Riemann tensor $R_{\mu \nu \alpha \beta}$. The $F(R, \mathcal{G})$ gravity, where $R$ and $\mathcal{G}$, respectively, denote the Ricci scalar and Gauss--Bonnet invariant, is another modified gravity theory that includes both the Ricci and Gauss--Bonnet scalars \cite{Elizalde10,Dombriz12}. The motivation behind this new gravity was to justify the evolution of the universe in the context of the dark energy and initial singularity.  The action for $F(R, \mathcal{G})$ gravity is
\begin{equation}\label{1}
S=\int \sqrt{-g} \frac{1}{2\kappa^2}F(R, \mathcal{G}) d^{4}x +\int \sqrt{-g} \mathcal{L}_m d^{4}x
\end{equation}
where $\mathcal{L}_m$ represents the matter Lagrangian and $\kappa^2=8 \pi G_{N}$, where $G_{N}$ is the Newtonian constant. The Gauss--Bonnet invariant can be expressed as $\mathcal{G} \equiv R^2-4R^{\mu \nu} R_{\mu \nu}+R^{\mu \nu \alpha \beta}R_{\mu \nu \alpha \beta}$. Now, by varying the action in Equation \eqref{1} with respect to the metric tensor $g_{\mu \nu}$, the field equations of $F(R, \mathcal{G})$ gravity can be expressed as
\begin{eqnarray} \label{2}
F_R{G}_{\mu\nu}&=&\kappa^2 T_{\mu\nu} + \frac{1}{2}g_{\mu\nu}[F(R, \mathcal{G})-RF_{R}] + \nabla_{\mu}\nabla_{\nu} F_{R} - g_{\mu\nu} \Box F_{R}\nonumber \\
&+&F_\mathcal{G}({-2R}{R_{\mu\nu}} + 4R_{\mu k}R^{k}_{\nu}-2R^{klm}_{\mu}R_{\nu k l m} + 4 g^{kl} g^{mn} R_{\mu k \nu m} R_{ln})\nonumber \\
&+&2(\nabla_{\mu}\nabla_{\nu}F_\mathcal{G})R - 2g_{\mu \nu}(\Box F_\mathcal{G})R+4(\Box F_\mathcal{G})R_{\mu\nu}-4(\nabla_{k} \nabla_{\mu} F_\mathcal{G})R^{k}_{\nu} \nonumber \\
&-&4(\nabla_{k} \nabla_{\nu} F_\mathcal{G})R^{k}_{\mu} + 4g_{\mu \nu}(\nabla_{k} \nabla_{l} F_\mathcal{G})R^{kl} - 4(\nabla_{l} \nabla_{n} F_\mathcal{G})g^{kl}g^{mn}R_{\mu k \nu m}
\end{eqnarray}
where $g_{\mu \nu}$, $G_{\mu \nu}$, and  $\nabla_{\mu}$, respectively, represent the gravitational metric potential, Einstein tensor, and covariant derivative operator associated with $g_{\mu \nu}$. The covariant d'Alembert operator, $\Box \equiv g^{\mu \nu}\nabla_{\mu}\nabla_{\nu}$ and ${T}_{\mu\nu}$ be the ordinary matter. We denote the partial derivative of $F(R, \mathcal{G})$ with respect to $R$ and $\mathcal{G}$, respectively, as, 
\begin{equation*}
F_{R} \equiv \frac{\partial F(R, \mathcal{G})}{\partial R},\hspace{1cm} F_{\mathcal{G}}\equiv \frac{\partial F(R, \mathcal{G})}{\partial \mathcal{G}}
\end{equation*}

{We obtain the following equation by taking the traces of Equation \eqref{2}
\begin{equation}
3 \Box F_R+ R F_R-2 F(R,\mathcal{G})+R (\Box F_{\mathcal{G}}+2 \frac{\mathcal{G}}{R} F_{\mathcal{G}})=\kappa^2 T
\end{equation}
where $T$ is trace of energy-momentum tensor and it is $T=g^{\mu \nu} T_{\mu \nu}=-\rho+3p$}.

Cognola et al. \cite{Cognola06} have studied the late-time acceleration issue in modified Gauss--Bonnet gravity and shown that this kind of model can experience the transition from deceleration to acceleration. In Refs. \cite{Felice09, Felice10}, the stability criteria for scalar and tensor perturbations in $F(R, \mathcal{G})$ gravity were investigated. Furthermore, Felice et al. \cite{Felice11} have derived the propagation speed for both odd-type and even-type perturbations in $F(R, \mathcal{G})$ gravity. Makarenko et al. \cite{Makarenko13} developed a phantom-type cosmological model in $F(R, \mathcal{G})$ gravity that does not lead to a future singularity. Laurentis et al. {\cite{Laurentis15}} have studied cosmological inflation and indicated the emergence of two double inflationary scenarios. \mbox{Martino et al.~{\cite{Martino20}}}  traced the cosmic history in $F(R, \mathcal{G})$ gravity and showed that it could lead gravity from ultraviolet to infrared scales. 
We consider the isotropic flat FLRW metric as 
\begin{equation} \label{3}
ds^{2}=-dt^{2}+a^{2}(t)(dx^{2}+dy^{2}+dz^{2}),
\end{equation}
where the scale factor $a(t)$ measures the expansion rate of the universe, and it appears that the expansion becomes uniform in the spatial directions. Using Equation \eqref{3}, the Ricci scalar $R$ and Gauss--Bonnet term $\mathcal{G}$ can be expressed in Hubble term, $H=\frac{\dot{a}}{a}$, as
\begin{equation} \label{4}
R=6 \left(\frac{\ddot{a}}{a}+\frac{\dot{a}^2}{a^2}\right)=6(\dot{H}+2H^{2}) \hspace{2cm} \mathcal{G}=24 \frac{{\ddot{a}}{\dot{a}^2}}{a^3}=24H^{2}(\dot{H}+H^{2})
\end{equation}

We consider the energy-momentum tensor in the form of perfect fluid as
\begin{equation} \label{5}
T_{\mu \nu}=(\rho+p)u_{\mu}u_{\nu}+pg_{\mu\nu} 
\end{equation}
where $\rho$ and $p$ are, respectively, the matter energy density and pressure. The observer $u_{\mu}$ with $u^{\mu}$ is the time-like four-velocity vector of the cosmic fluid satisfying $u_{\mu} u^{\mu} =-1$ for the space--time Equation \eqref{3} and energy-momentum tensor Equation \eqref{5}. The $F(R,\mathcal{G})$ gravity field equations Equation \eqref{2} can be obtained as 
\begin{equation}
3H^{2}F_{R}=\kappa \rho+\frac{1}{2}[RF_{R}+\mathcal{G} F_\mathcal{G}-F(R, \mathcal{G})]-3H \dot{F}_{R}-12H^{3}\dot{F}_\mathcal{G}, \label{6}
\end{equation}
\begin{equation}
2\dot{H}F_{R}+3H^{2}F_{R}=-\kappa p+\frac{1}{2}\left[RF_{R}+\mathcal{G} F_\mathcal{G}-F(R, \mathcal{G})\right]-2H\dot{F}_{R}-\ddot{F}_{R}-4H^{2}\ddot{F}_\mathcal{G}-8H\dot{H}\dot{F}_\mathcal{G}-8H^{3}\dot{F}_\mathcal{G} \label{7}\\
\end{equation}

An over-dot represents an ordinary derivative with respect to cosmic time $t$.  The energy density and matter pressure can be obtained if the functional $F(R,\mathcal{G})$ has some explicit form. Thus, here we have assumed the quadratic form of $F(R,\mathcal{G})$ \cite{Laurentis15} as,
\begin{equation} \label{8}
F(R, \mathcal{G})=R+\alpha R^{2}+\beta \mathcal{G}^{2},
\end{equation}
where $\alpha$ and $\beta$ are pairing constants. {{The linear component in $F(R, \mathcal{G})$ is included to generate the correct weak field limit. We analyzed the $R^2$ model with a correction that introduces extra degrees of freedom owing to the inclusion of the Gauss--Bonnet component. Because the linear one does not contribute, the term $\mathcal{G}^2$ is the first important term in $\mathcal{G}$ in the above Lagrangian.}}

The dynamical parameter's behavior can be analyzed if we express these in terms of cosmic time. Since our objective is to avoid the initial singularity, we need to constrain the Hubble parameter that supports bouncing behavior. In bouncing cosmology: (i) At the bouncing epoch, the Hubble parameter vanishes, the deceleration parameter becomes singular, and the scale factor contracts to a non-zero finite value; (ii) From the bouncing point, the Hubble parameter changes sign, 
and the null energy conditions are violated; (iii) After the bounce, the slope of the scale factor increases and the Hubble parameter, respectively, is negative and positive during matter contraction and expansion phases
. Keeping these properties of bouncing cosmology in mind, we consider two scale factors in the following sections that support these behaviors.

\section{ Bouncing Model I} \label{SEC III}
In this section, we consider the scale factor as $a(t)=\left(1+\frac{3}{4}{\chi t^2}\right)^\frac{1}{3}$, where $\chi$ is the scale factor parameter and can be constrained on an observational and physical basis \cite{Haro14, Haro12}. The corresponding Hubble parameter that measures the rate of expansion of the universe and the deceleration parameter that decides the accelerating or decelerating behavior of the universe can be, respectively, given as
\begin{eqnarray}
H(t)=\frac{2 \chi t}{3 \chi t^2+4}, \hspace{1cm}
q=\frac{1}{2}-\frac{2}{\chi t^2} \label{11}
\end{eqnarray}

\begin{figure} [H]
\centering
\includegraphics[width=65mm]{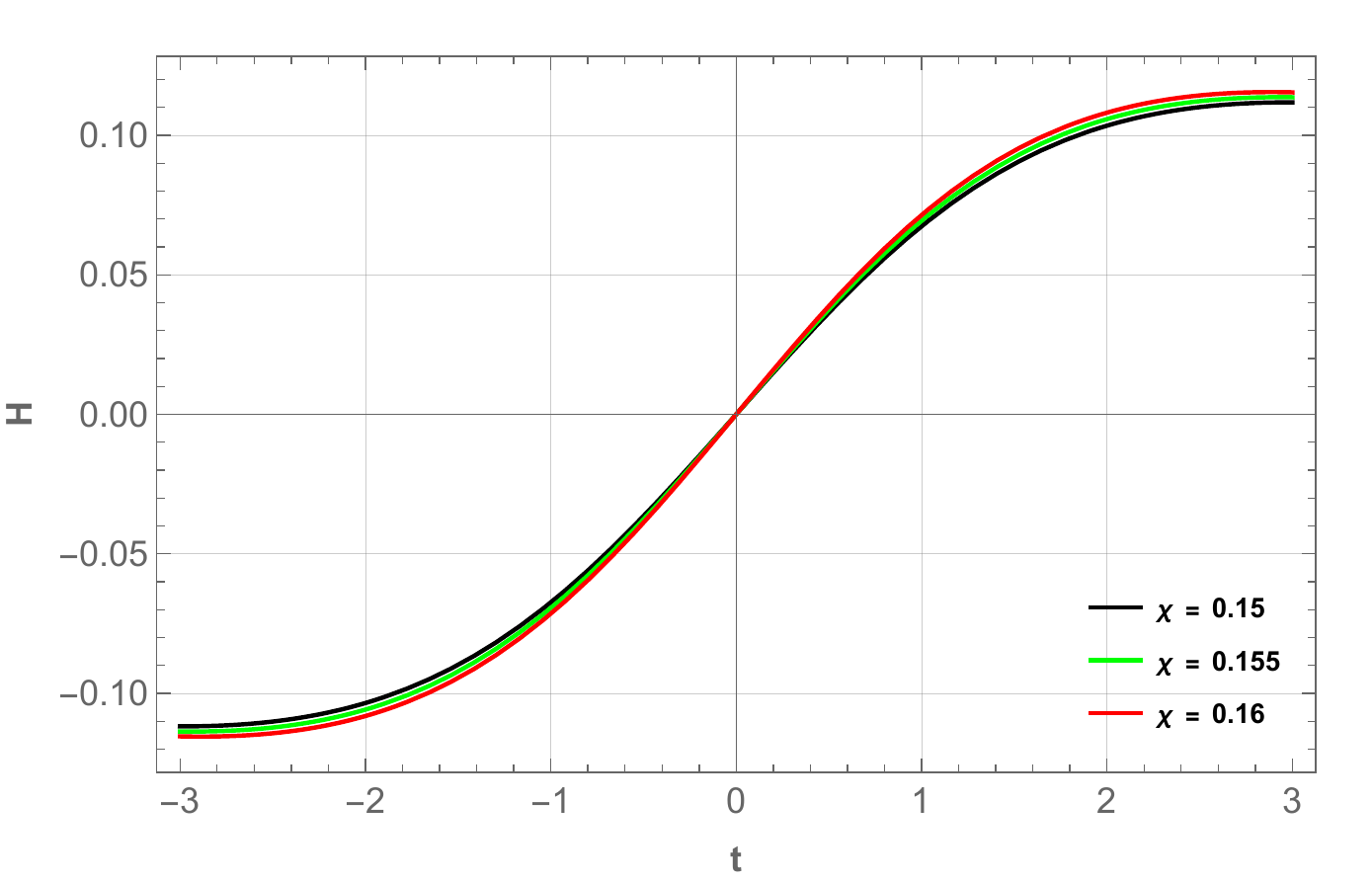}
\includegraphics[width=65mm]{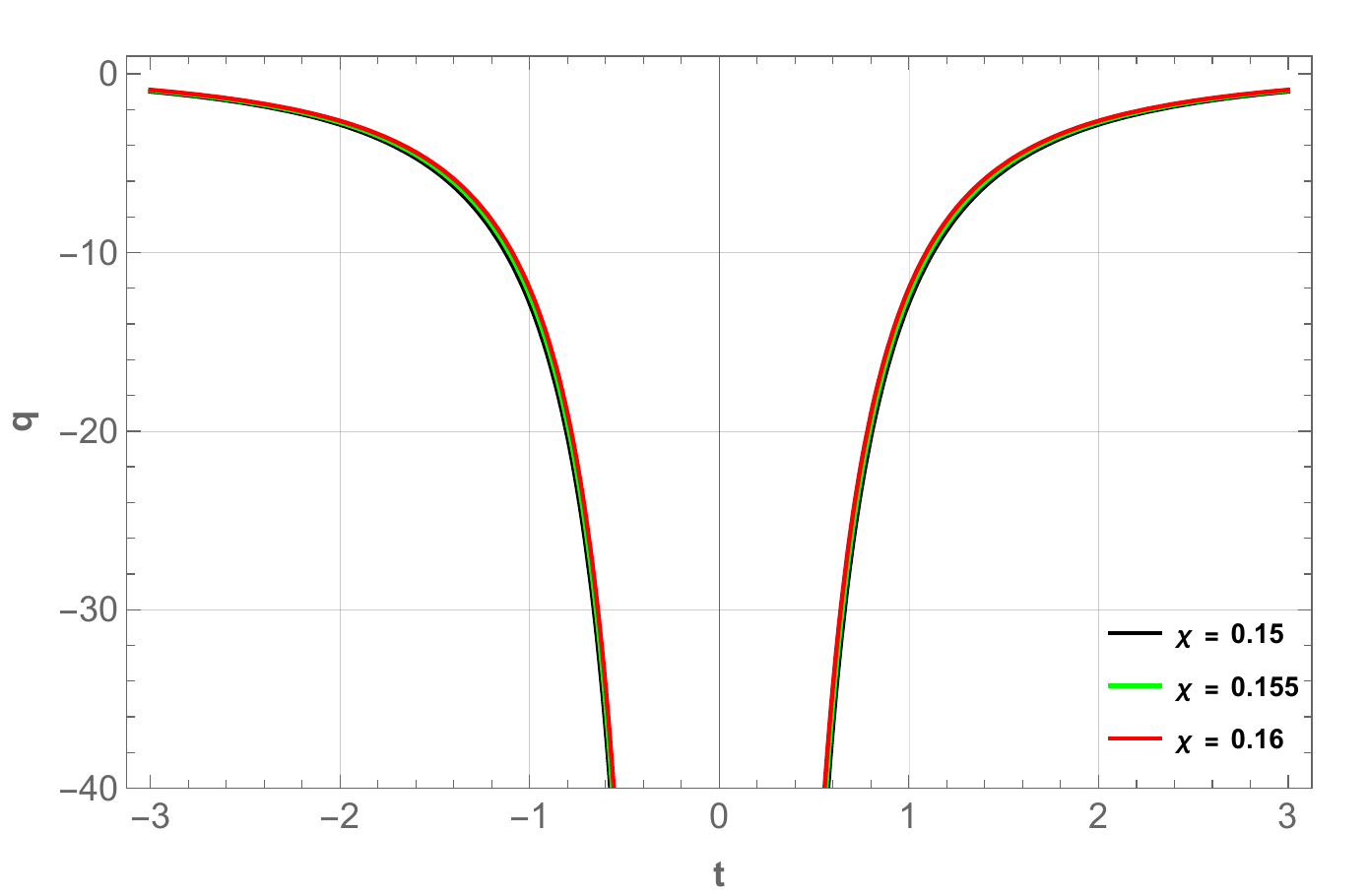}
\caption{Behavior of Hubble parameter (\textbf{left panel}) and deceleration parameter (\textbf{right panel}) versus cosmic time $t$.
} \label{FIG1}
\end{figure}

The Hubble parameter increases from the early universe and vanishes at the epoch $t=0$; then, it further increases, as expected in the bouncing behavior (Figure \ref{FIG1} (left panel)). A negative or positive value of $q$ signifies, respectively, the accelerating and decelerating behavior of the universe. In Figure \ref{FIG1} (right panel), the time evolution of the deceleration parameter, which is independent of the model parameter, has been shown for three representative values of the scale factor parameter $\chi$. The deceleration parameter is symmetric with regard to the bouncing time $t=0$. If $\chi > 0$, $q$ becomes negative and hence shows the accelerating behavior, i.e., $\chi t^2 < 4$.

Substituting the Hubble parameter Equation \eqref{11}, the dynamical parameters Equations \eqref{6} and \eqref{7} can be expressed as
\begin{eqnarray} 
\rho&=&\frac{1}{\eta^8}\big({12 \chi^2 \eta^4 (t^2 \eta^2-54 \alpha  \chi t^2 (\chi t^2+8)-96 \alpha )+18,432 \beta  \chi^6 t^4 (\chi t^2 (23 \chi t^2-168)+48)}\big),\nonumber\\ \label{12}
p&=&\frac{1}{\eta^8}\Big(6144 \beta  \chi^5 t^2 \big(\chi^2 t^4 (207 \chi t^2-2384)+2736 \chi t^2-384\big) -16 \chi \eta^6\Big) \nonumber\\& & -\frac{216 \alpha \chi^2}{\eta^4} \Big(\chi t^2 (3 \chi t^2+40)-16\Big)
\end{eqnarray}

\begin{figure} [H]
\centering
\includegraphics[width=65mm]{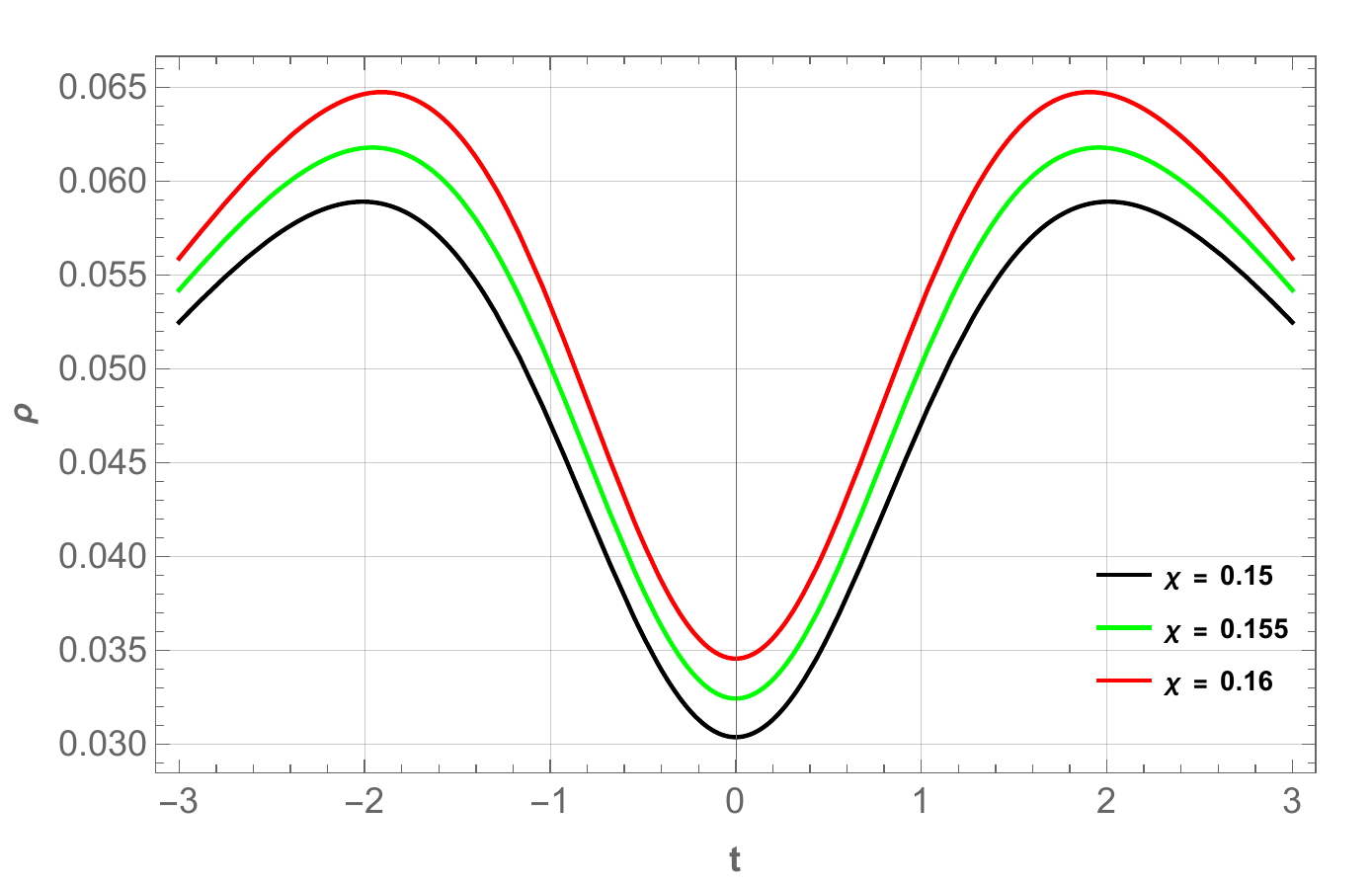}
\includegraphics[width=65mm]{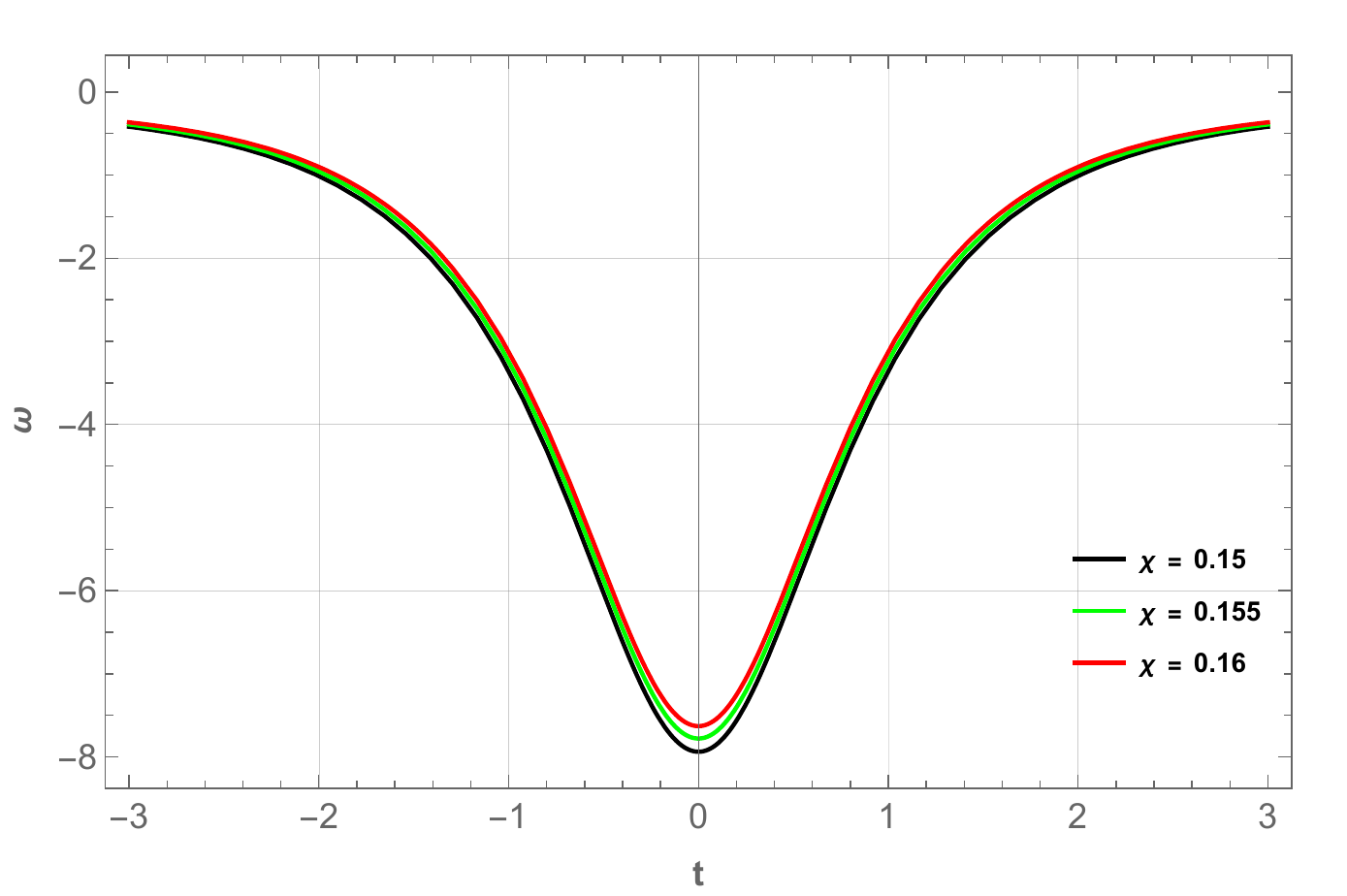}
\caption{Behavior of energy density (\textbf{left panel}) and EoS parameter (\textbf{right panel}) versus cosmic time $t$ for $\alpha=-0.30, \beta=0.15$.}
\label{FIG2}
\end{figure}

The equation of state (EoS) parameter $\omega=\frac{p}{\rho}$ can be obtained from Equation \eqref{12} as
\begin{eqnarray}
\omega &=\frac{6144 \beta  \chi^5 t^2 [\chi t^2 \left(\chi t^2 (207 \chi t^2-2384)+2736\right)-384] + 8 \chi \eta^4 [-27 \alpha  \chi \left(\chi t^2 (3 \chi t^2+40) - 16\right)-2 \eta^2]}{12 \chi^2 \eta^4 [t^2 \eta^2-54 \alpha  \chi t^2 (\chi t^2+8) - 96 \alpha ]+18,432 \beta  \chi^6 t^4 [\chi t^2 (23 \chi t^2-168)+48]}, \label{14}
\end{eqnarray}
where $\eta=3\chi t^{2}+4$. In Figure \ref{FIG2} (left panel), the behavior of energy density has been shown with three representative values, namely $\chi= 0.15, 0.155,0.16$. There is a sharp increase in energy density immediately after the bounce, and after some time, it starts decreasing. The behavior remains the same both before and after the bounce. The EoS parameter Figure~\ref{FIG2}~(right panel) remains mostly in the phantom region in the time range of $t\in(-2,2)$. When the bouncing epoch occurs, the acceleration should be such that the Hubble parameter $\dot{H}>0$ is satisfied, which favors the phantom behavior. The EoS parameter for a bouncing model evolves in the phantom zone and crosses the phantom split $\omega=-1$ at least twice, once before and once after the bounce \cite{Cai10}. When the dynamical aspect of the EoS parameter has been studied, it evolves from a lower negative value, crosses the phantom split, and reaches a minimum in the phantom-like region at the bouncing epoch. The same behavior can be observed in Figure \ref{FIG2} (right panel).

The phantom $\omega <-1$ \cite{Caldwell02}, quintessence $-1<\omega\leq 0$ \cite{Steinhardt99}, and the quintom $\omega$ cross $-1$ are three prominent classes of scalar-field dark-energy models accessible in the literature to explore the dark energy features of the model. The quintom scenario can be used to travel from the phantom to the quintessence realm. The model can transition from $\omega>-1$ to $\omega<-1$ since the EoS value is time-dependent \cite{Wang06}.

The energy conditions assign the underlying causal and geodesic structure of space--time. Any geometrical extension of GR has to deal with the standard energy conditions~\cite{Nojiri11,Capozziello11}. Violating certain energy conditions is essential in the extended theories and bouncing cosmology. Hence, we present here null energy condition (NEC)---$ \rho+p \geq 0$; a weak energy condition (WEC)---$\rho \geq 0$; $\rho + p \geq 0$; a strong energy condition (SEC)---$ \rho+3p \geq 0$; and a dominant energy condition (DEC)---$\rho-p \geq 0$. These energy conditions are not independent since WEC $\Rightarrow$ NEC, SEC $\Rightarrow$ NEC, DEC $\Rightarrow $WEC, {as we notice that all other \textit{pointwise} energy conditions will also be violated if the NEC is violated \cite{Novello08}}. For Model I, the energy conditions can be derived using Equations \eqref{6} and \eqref{7} as
\begin{eqnarray}
&\rho+p = \frac{4 \chi \eta^4 \left[\left(3 \chi t^2-4\right) \eta^2-36 \alpha  \chi \left(9 \chi^2 t^4+96 \chi t^2-16\right)\right]+24,576 \beta  \chi^5 t^2 \left[\chi t^2 \left(\chi t^2 \left(69 \chi t^2-722\right)+720\right)-96\right]}{\eta^8},\\
&\rho+3p =\frac{36,864 \beta  \chi^5 t^2 \left[\chi t^2 \left(\chi t^2 \left(115 \chi t^2-1276\right)+1392\right)-192\right]+12 \chi \eta^4 \left[\left(\chi t^2-4\right) \eta^2-24 \alpha  \chi \left(9 \chi t^2 \left(\chi t^2+12\right)-32\right)\right]}{\eta^8},\\
&\rho-p =\frac{4 \chi \eta^4 \left[288 \alpha  \chi \left(3 \chi t^2-4\right)+\eta^3\right]-12,288 \chi^5 \beta  t^2 \left[\chi t^2 \left(\chi t^2 \left(69 \chi t^2-940\right)+1296\right)-192\right]}{\eta^8} \label{15}
\end{eqnarray}

\begin{figure} [H]
\centering
\includegraphics[width=65mm]{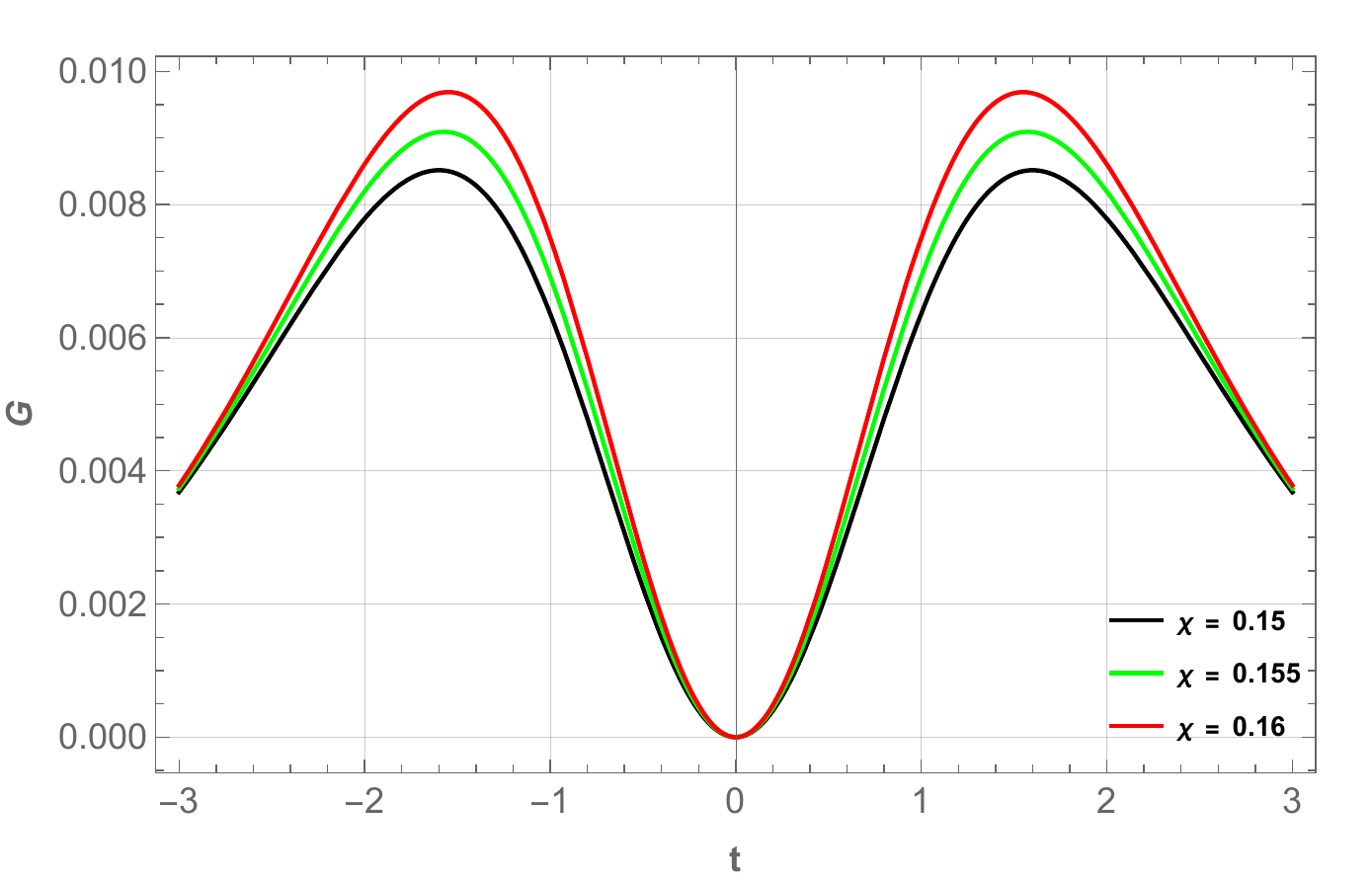}
\includegraphics[width=65mm]{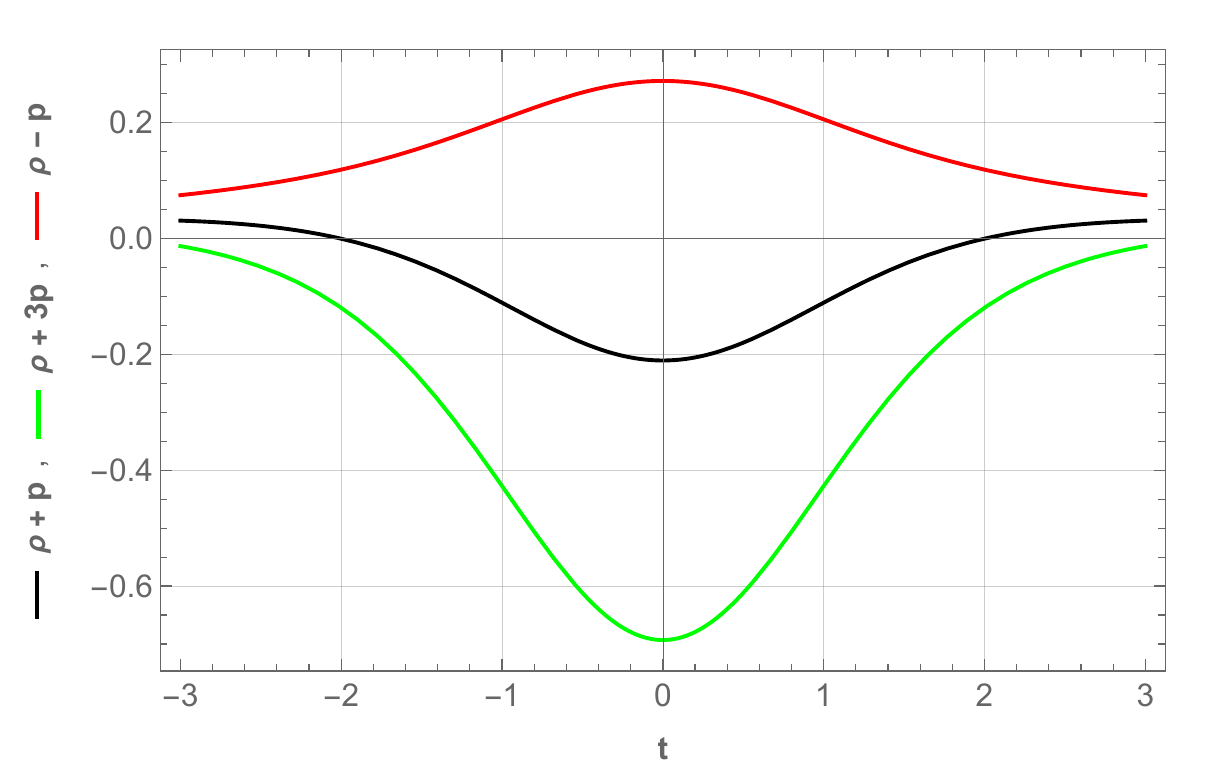}
\caption{Behavior of Gauss--Bonnet invariant (\textbf{left panel}) and energy conditions (\textbf{right panel}) versus cosmic time $t$. The parameters values are $\alpha = -0.30$, $\beta = 0.15$.}
\label{FIG3}
\end{figure}

The Gauss--Bonnet invariant is an important component in this extended gravity, which can also be expressed in terms of the Hubble parameter. Since we are investigating the model with a known scale factor, the behavior of $\mathcal{G}$ must also be observed. Figure~\ref{FIG3} (left panel) shows the bouncing behavior of the Gauss--Bonnet invariant, which entirely remains in the positive domain. It is symmetrical and reaches a peak immediately after the bounce in both positive and negative time zones. Subsequently, it decreases after attaining the peak at $t \approx \pm 1.55$.  The behavior of energy conditions shows the violation of $\rho+3p$ and $\rho+p$ at the bounce; however, it satisfies the $\rho-p$. The $\rho+p$ decreased and kept falling to a negative value in the negative cosmic time domain and increased from negative values in the positive cosmic time domain. At the bounce epoch, the null energy condition is violated, and the apparent phenomenon means that the strong energy conditions are also violated; nonetheless, it has been observed that it is violated throughout the universe's evolution. Hence, the model validates the bouncing behavior in $F(R,\mathcal{G})$ gravity.

In any theory of gravity, the geometrical parameters play a major role in analyzing the model. Apart from Hubble and deceleration parameters, there are other geometrical parameters such as the jerk parameter ($j$) and snap parameter ($s$).  The jerk parameter decides the rate of acceleration change, and the snap parameter measures the jerk rate. The deceleration parameter is insufficient to account for the entirety of cosmic dynamics; hence, the sign of $j$ is vital to observe whether the universe changes during evolution. To note, a positive jerk parameter indicates the change in the expansion of the universe at some point in its evolution. The ($j,s$) pair known as the state finder pair \cite{Sahni03,Alam03} can be used to distinguish between different dark energy models and can be obtained for the bouncing scale factor as
\begin{eqnarray}
j&=&q+2q^2-\frac{\dot{q}}{H}=1-\frac{12}{\chi t^2}, \nonumber \\
s&=&\frac{j-1}{3(q-\frac{1}{2})}=\frac{12 \left(7 \chi t^2-2\right)}{\chi^2 t^4}-\frac{7}{2} \label{16}
\end{eqnarray}

\begin{figure} [H]
\centering
\includegraphics[width=65mm]{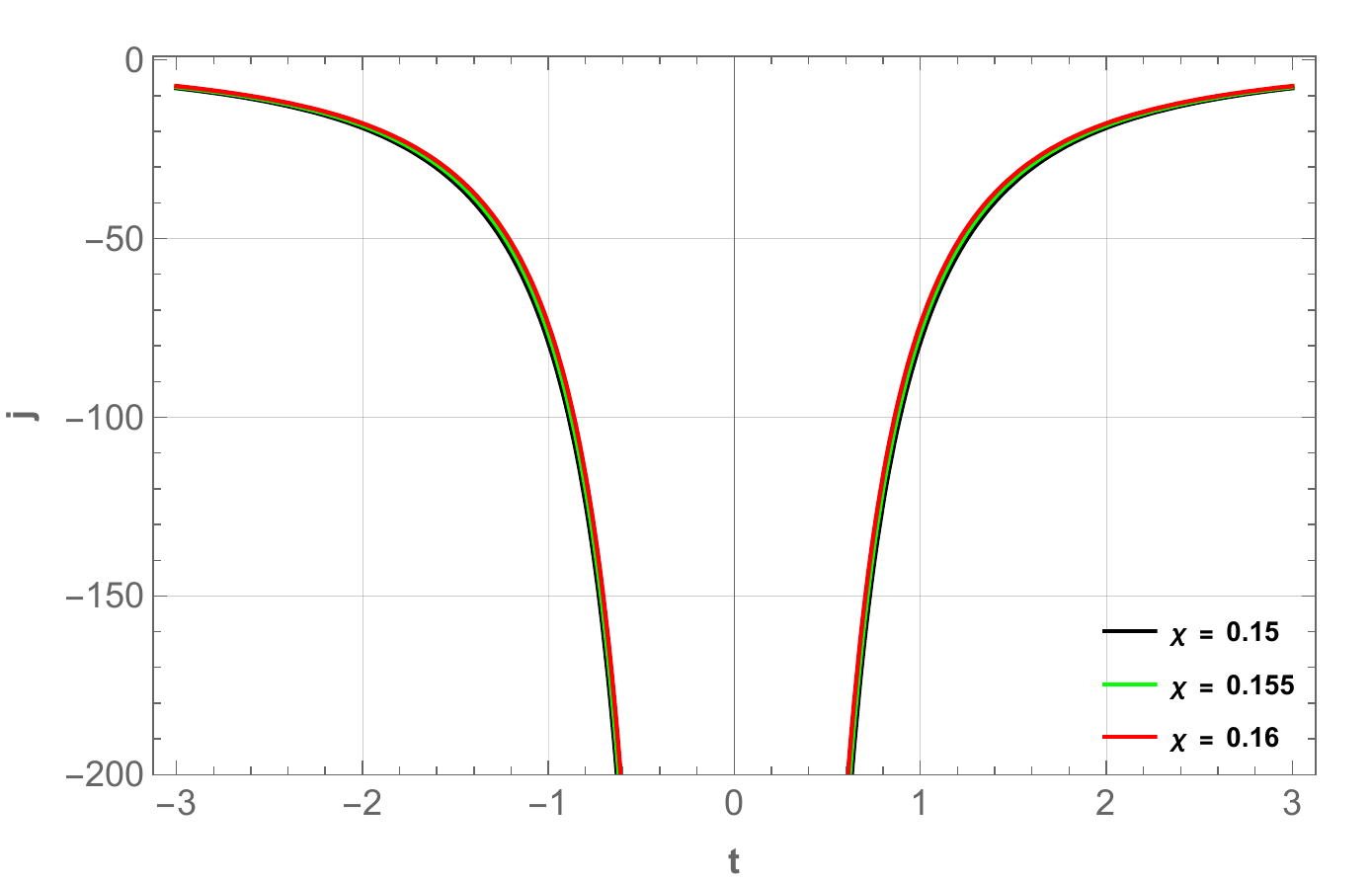}
\includegraphics[width=65mm]{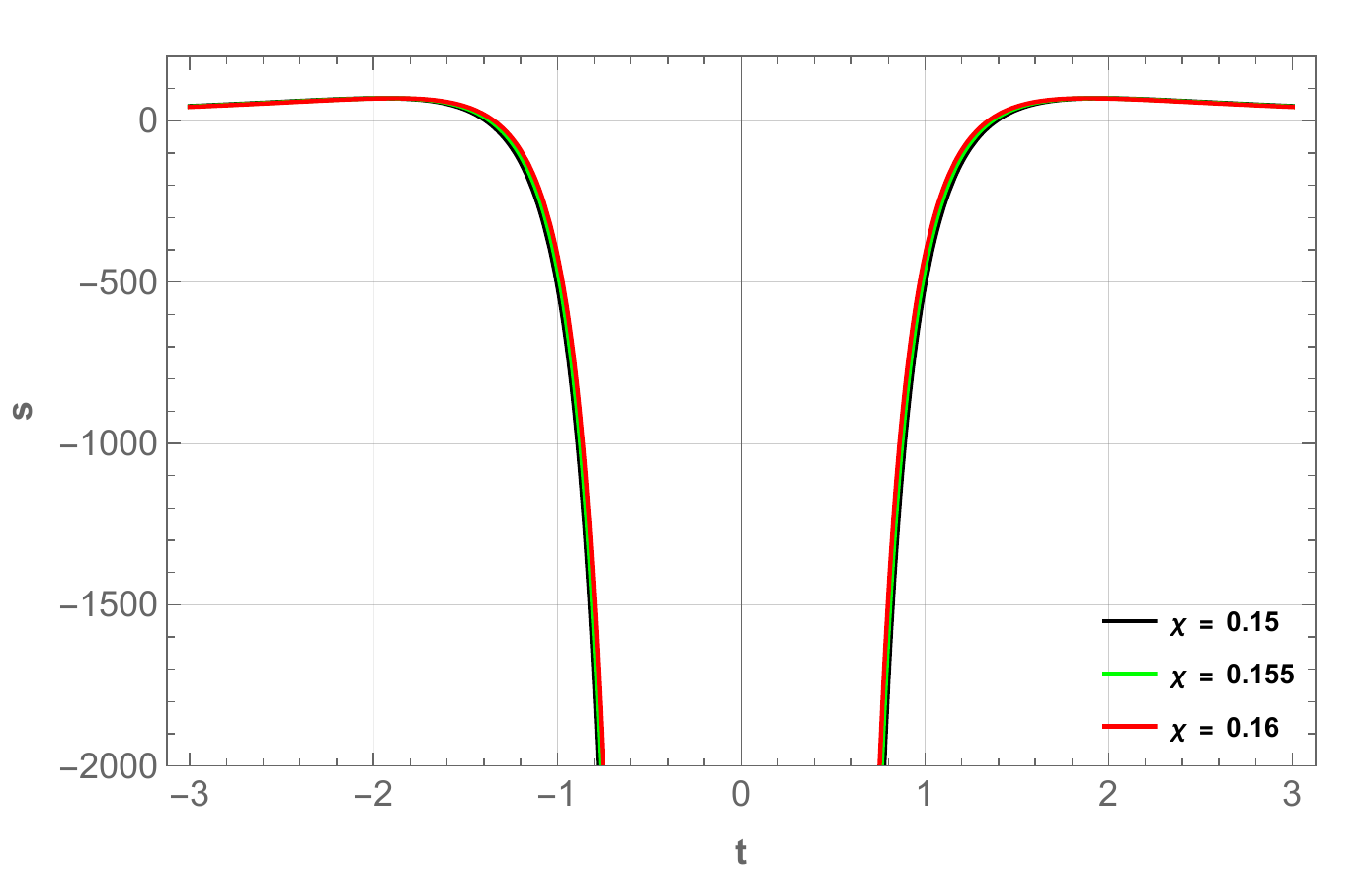}\\
\includegraphics[width=62mm]{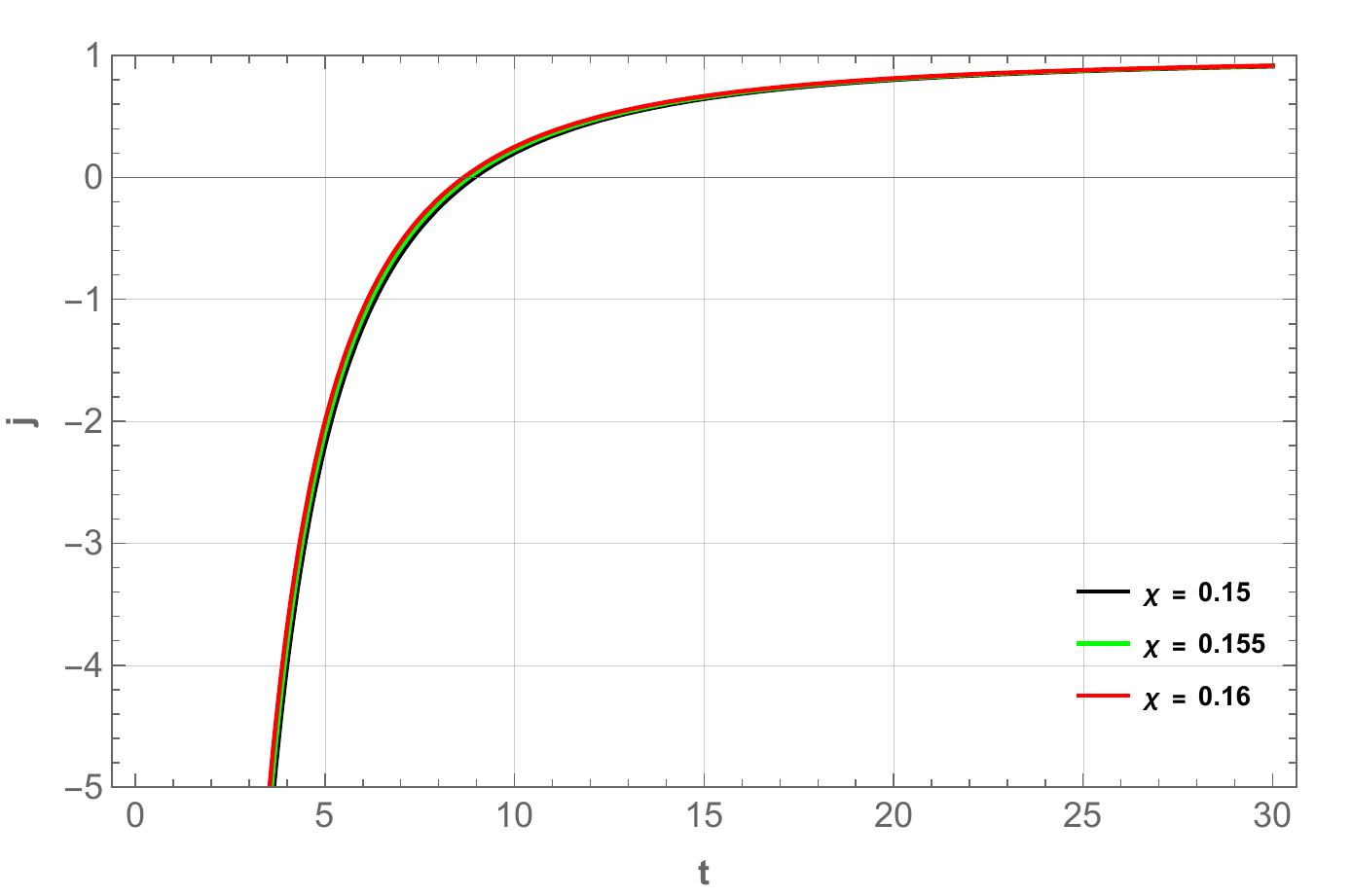}
\includegraphics[width=65mm]{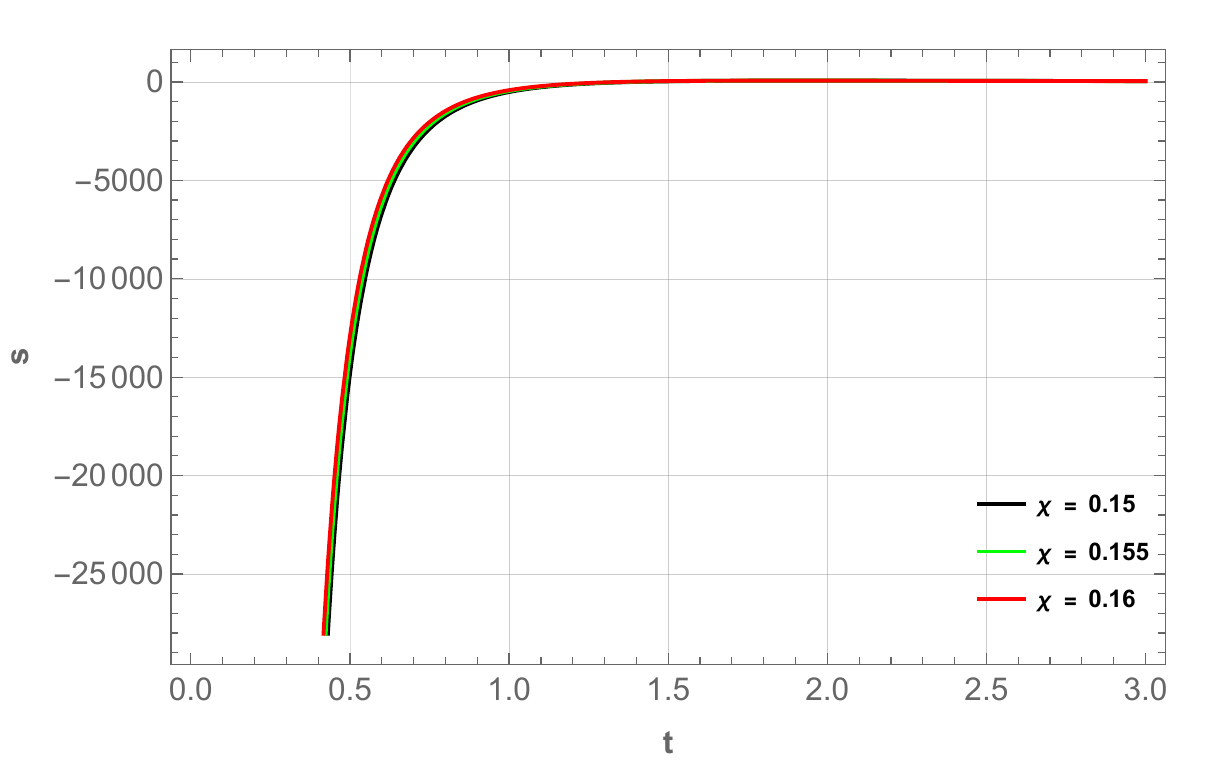}
\caption{Symmetric behavior of the jerk parameter (\textbf{above left panel}) and snap parameter (\textbf{above right panel}) versus cosmic time $t$. The jerk (\textbf{below left panel}) and snap (\textbf{below right panel}) parameter in positive time domain.}
\label{FIG4}
\end{figure}

In Figure \ref{FIG4}, both the jerk and snap parameters show similar behavior with singularity appearing at the bounce epoch $t=0$. In the negative and positive time zones, both are, respectively, showing decreasing and increasing behavior. Nonetheless, it evolves from the $\Lambda$CDM behavior $(j=1)$ in both the negative and positive time zones. A snap parameter, which is determined via the fourth derivative of the scale factor, exhibits the same behavior but departs from the mandated $\Lambda$CDM behavior $(s=0)$.

\section{Bouncing Model II} \label{SEC IV}
As a second model, we consider another bouncing scale factor in the form $a(t)=\left(\frac{\gamma}{\lambda}+t^2 \right)^\frac{1}{2\lambda}$, where $\gamma$ and $\lambda$ are the parameters of the scale factor, and on a physical basis, these can be constrained \cite{Abdussattar11}. The corresponding Hubble and deceleration parameters are
\begin{eqnarray}
H=\frac{t}{\gamma+t^2 \lambda}, \hspace{1cm}
q=-1+\lambda-\frac{\gamma}{t^2} \label{17}
\end{eqnarray}
\begin{figure} [H]
\centering
\includegraphics[width=65mm]{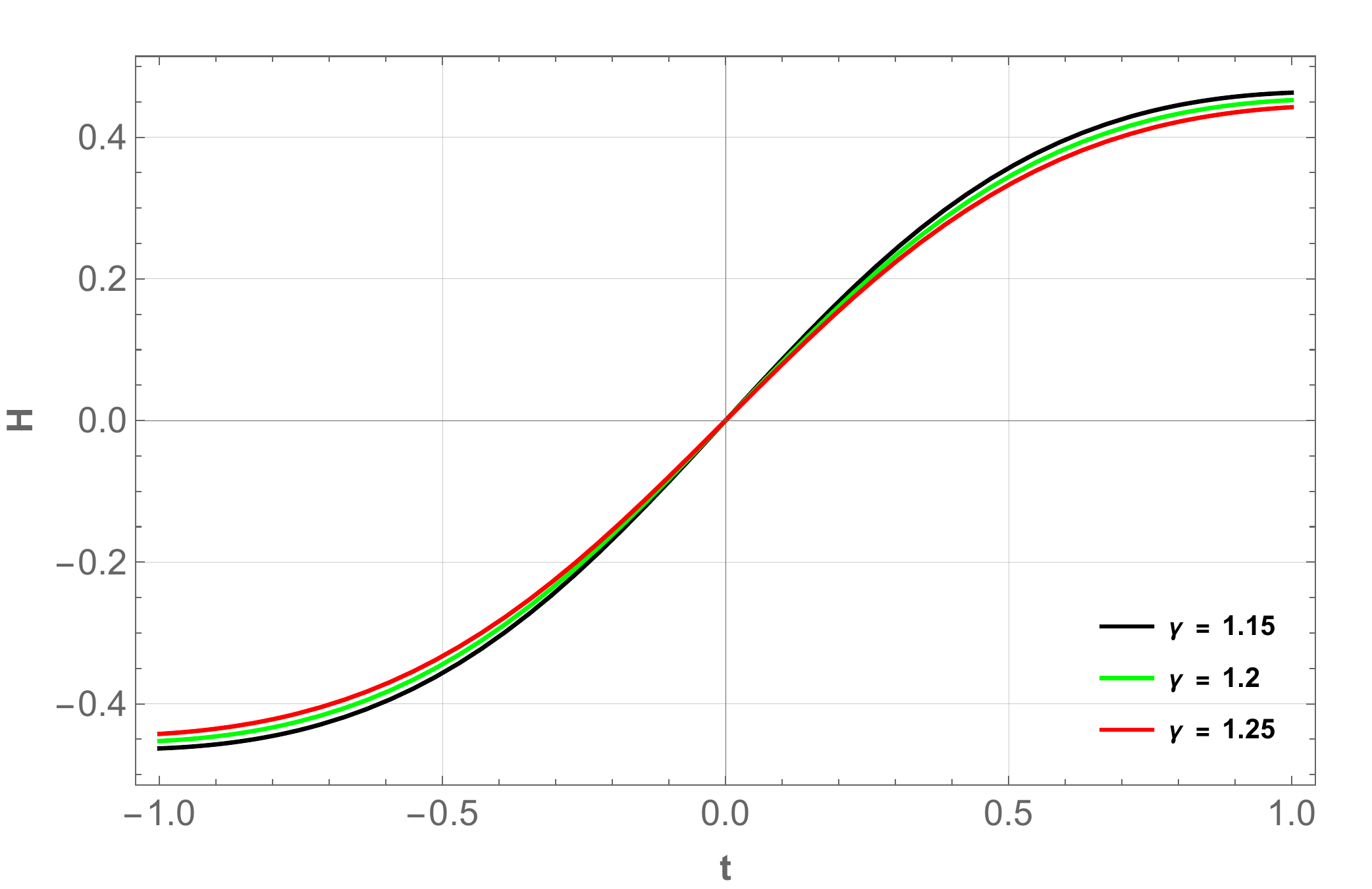}
\includegraphics[width=65mm]{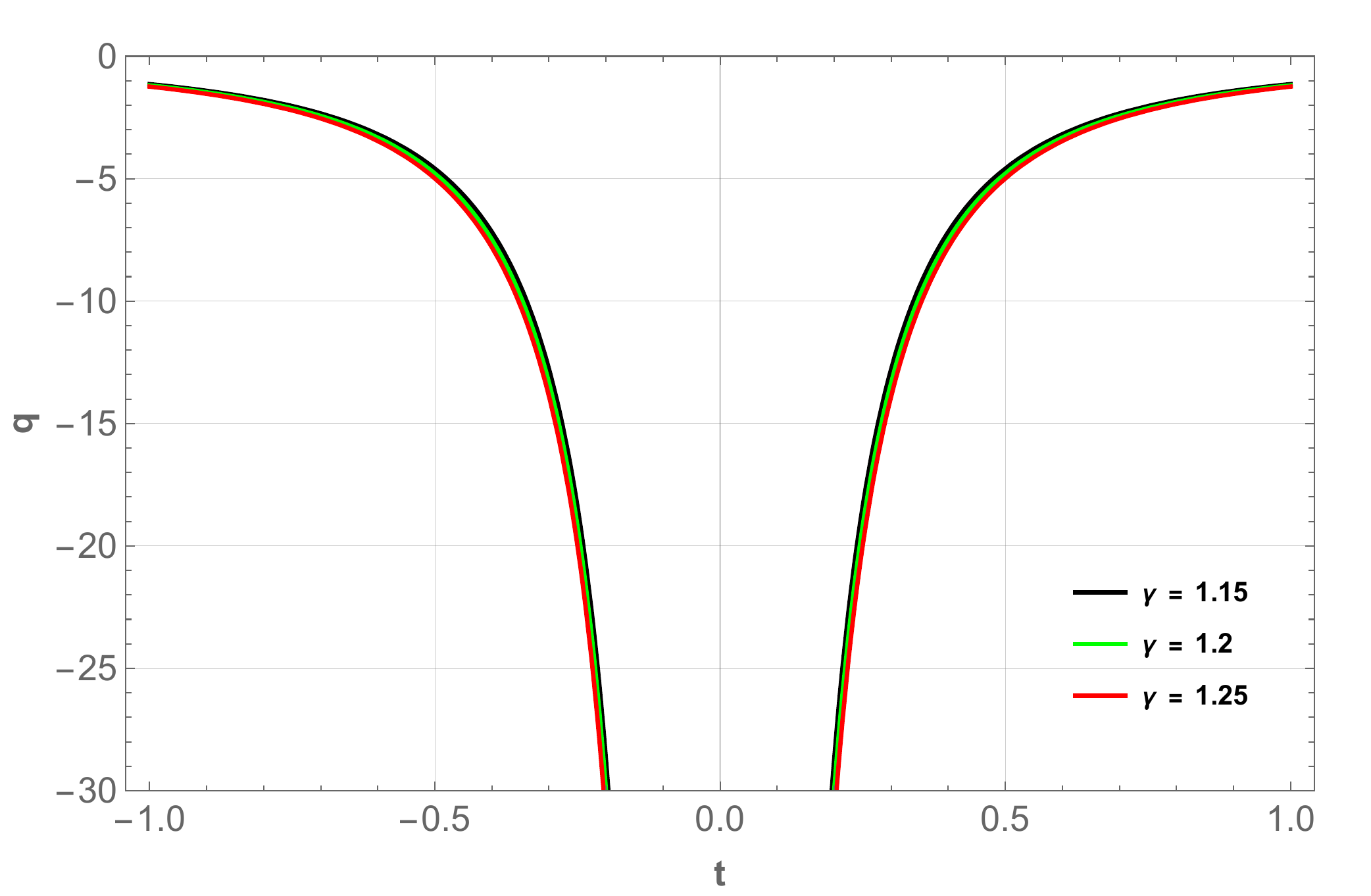}
\caption{Behavior of the Hubble parameter (\textbf{left panel}) and deceleration parameter (\textbf{right panel}) versus cosmic time $t$ with $\lambda=1.01$.} 
\label{FIG5}
\end{figure}
We can observe from Equation \eqref{17} that, at $t=0$, the Hubble parameter vanishes and the deceleration parameter becomes constant ($-1+\lambda$) at $t \rightarrow \infty$. Subsequently, the acceleration or deceleration behavior lies with the value of $\lambda$, i.e., for $\lambda<1$, the model accelerates and decelerates for $\lambda>1$. Furthermore, we can say that the acceleration behavior lies with the expression $(\lambda < \frac{\gamma}{t^2}+1)$, and the model decelerates for $(\lambda > \frac{\gamma}{t^2}+1)$. Figure \ref{FIG5} shows the graphical behavior of both parameters. Both parameters support the bouncing behavior. There are limited changes observed in the deceleration parameter with the representative values of the scale factor parameter $\gamma$. Now, with the form of the Hubble parameter Equation~\eqref{17}, the energy density and matter pressure can be calculated, respectively, from Equations \eqref{6} and \eqref{7} as
\begin{eqnarray} 
\rho&=&\frac{1}{\kappa^{2}}\big(3H^2+108\alpha \dot{H} H^2-18\alpha \dot{H}^2-288\beta H^8+1728\beta \dot{H} H^6+864\beta \dot{H}^2H^4+36\alpha H\ddot{H} \nonumber \\ & &+576\beta\ddot{H}H^5\big) \label{18}\\
\nonumber p&=&\frac{1}{\kappa^{2}}\big(-2\dot{H}-3H^2-54\alpha \dot{H}^2-108\alpha \dot{H} H^2+288\beta H^8-960\beta \dot{H} H^6-4320\beta \dot{H}^2H^4 \nonumber \\ & &-72\alpha H\ddot{H}-12\alpha \dot{\ddot{H}}-1152\beta \ddot{H}H^5-1152\beta H^2 \dot{H}^3-1536\beta \dot{H} \ddot{H} H^3-192\beta \dot{\ddot{H}}H^4 \big) \label{19}
\end{eqnarray}
where
\begin{eqnarray*}
\dot{H}&=&\frac{1}{\gamma+\text{t}^2 \lambda}-\frac{2 \text{t}^2 \lambda}{\left(\gamma+\text{t}^2 \lambda\right)^2},\\ \ddot{H}&=&\frac{8 \text{t}^3 \lambda^2}{\left(\gamma+\text{t}^2 \lambda\right)^3}-\frac{6 \text{t} \lambda}{\left(\gamma+\text{t}^2 \lambda\right)^2},\\ \dot{\ddot{H}}&=&-\frac{48 t^4 \lambda^3}{(\gamma + t^2 \lambda)^4} + \frac{48 t^2 \lambda^2}{(\gamma + t^2 \lambda)^3} - \frac{(6 \lambda)}{(\gamma + t^2 \lambda)^2}
 \end{eqnarray*}

From the above Equations \eqref{18} and \eqref{19}, the EoS parameter can be obtained as $\omega=\frac{p}{\rho}$

\begin{figure} [H]
\centering
\includegraphics[width=65mm]{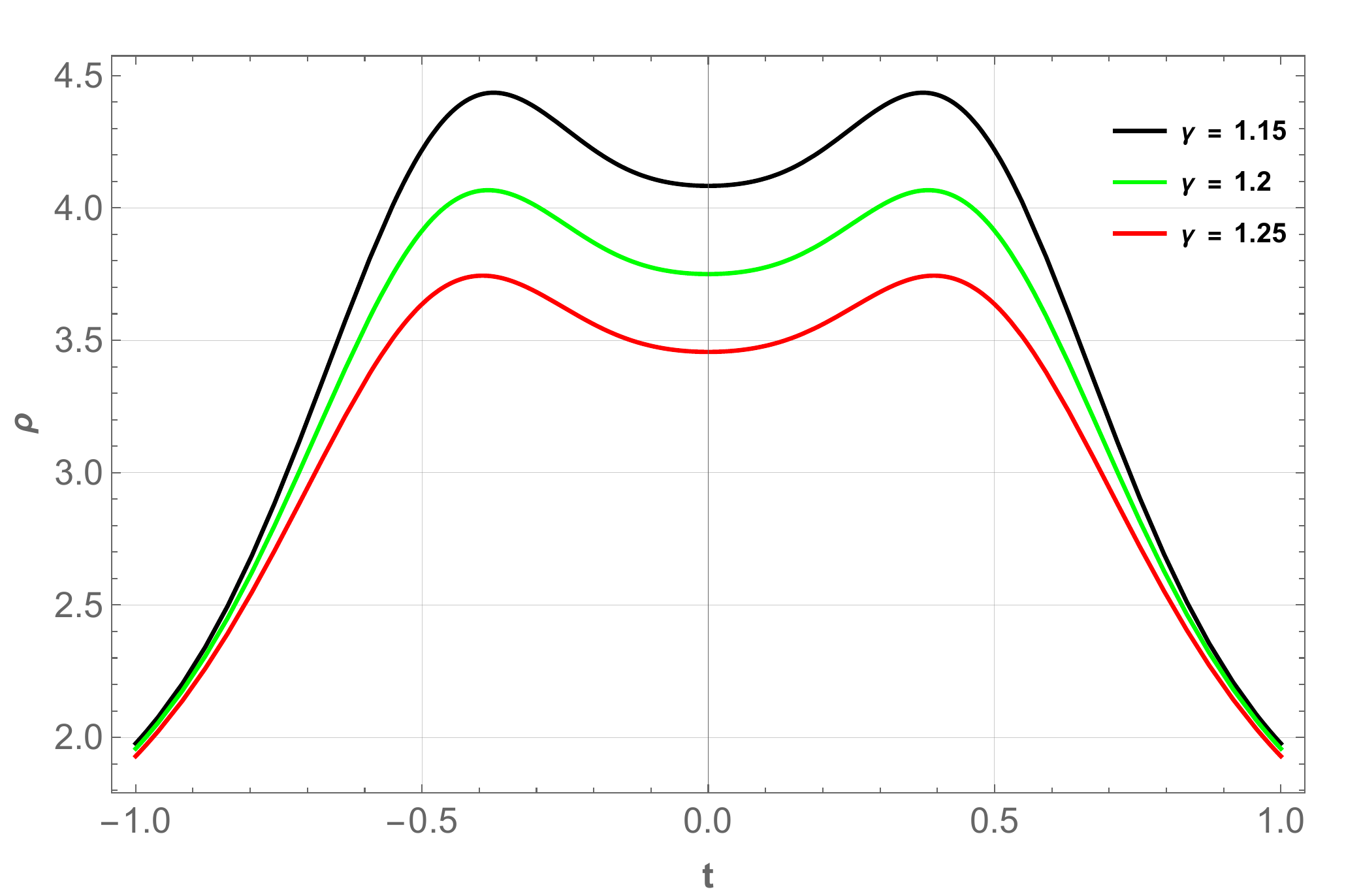}
\includegraphics[width=65mm]{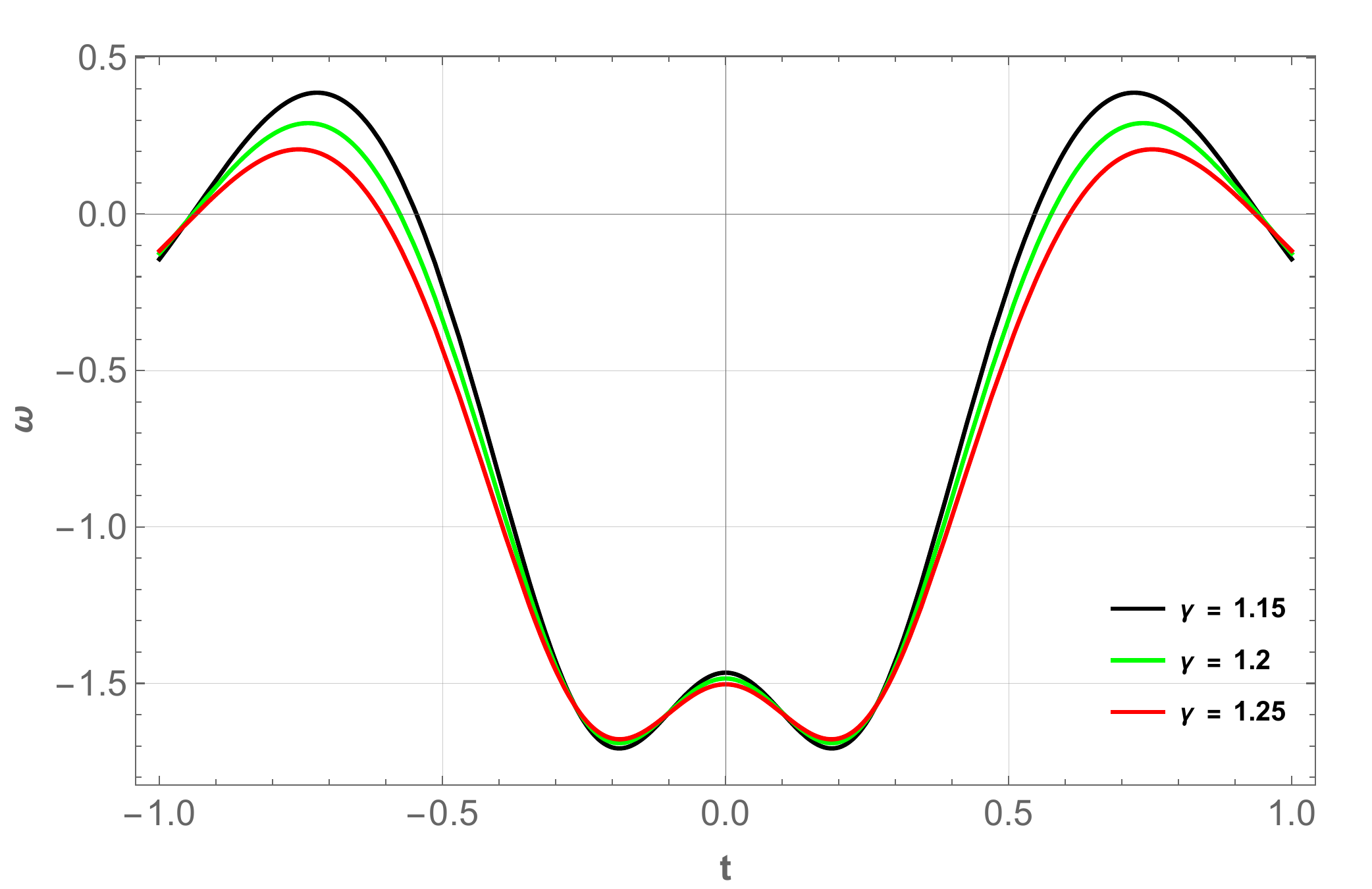}
\caption{Behavior of energy density (\textbf{left panel}) and EoS parameter (\textbf{right panel}) versus cosmic time $t$, $\lambda=1.01$.}
\label{FIG6}
\end{figure}

To maintain positive energy density (left panel) during the entire evolution, the minimum matter geometry coupling scale factor parameter $\gamma$ should be restricted to a positive value. The current model is depicted in Figure \ref{FIG6} for three representative values of the scale factor parameter, namely $\gamma = 1.15, 1.20, 1.25$. The phantom phase at $-0.39 < t < 0.39$ is shown in the EoS parameter (right panel). The energy density becomes a positive number for these values in both the positive and negative time domains. The bouncing epoch acceleration should be such that the Hubble parameter $\dot{H}>0$ is satisfied, favoring the phantom behavior of the model. Near the bounce, the EoS parameter is well shaped, and there are noticeable variations in the depth of the well as the value of $\gamma$ changes. The EoS parameter evolves from the phantom area on both sides of the bouncing epoch, and as it travels away from the bouncing era, the model passes through the CDM line and remains in the quintessence phase. In both the negative and positive time zones, similar behavior has been observed in the evolution of the EoS curve. 

\begin{figure} [H]
\centering
\includegraphics[width=65mm]{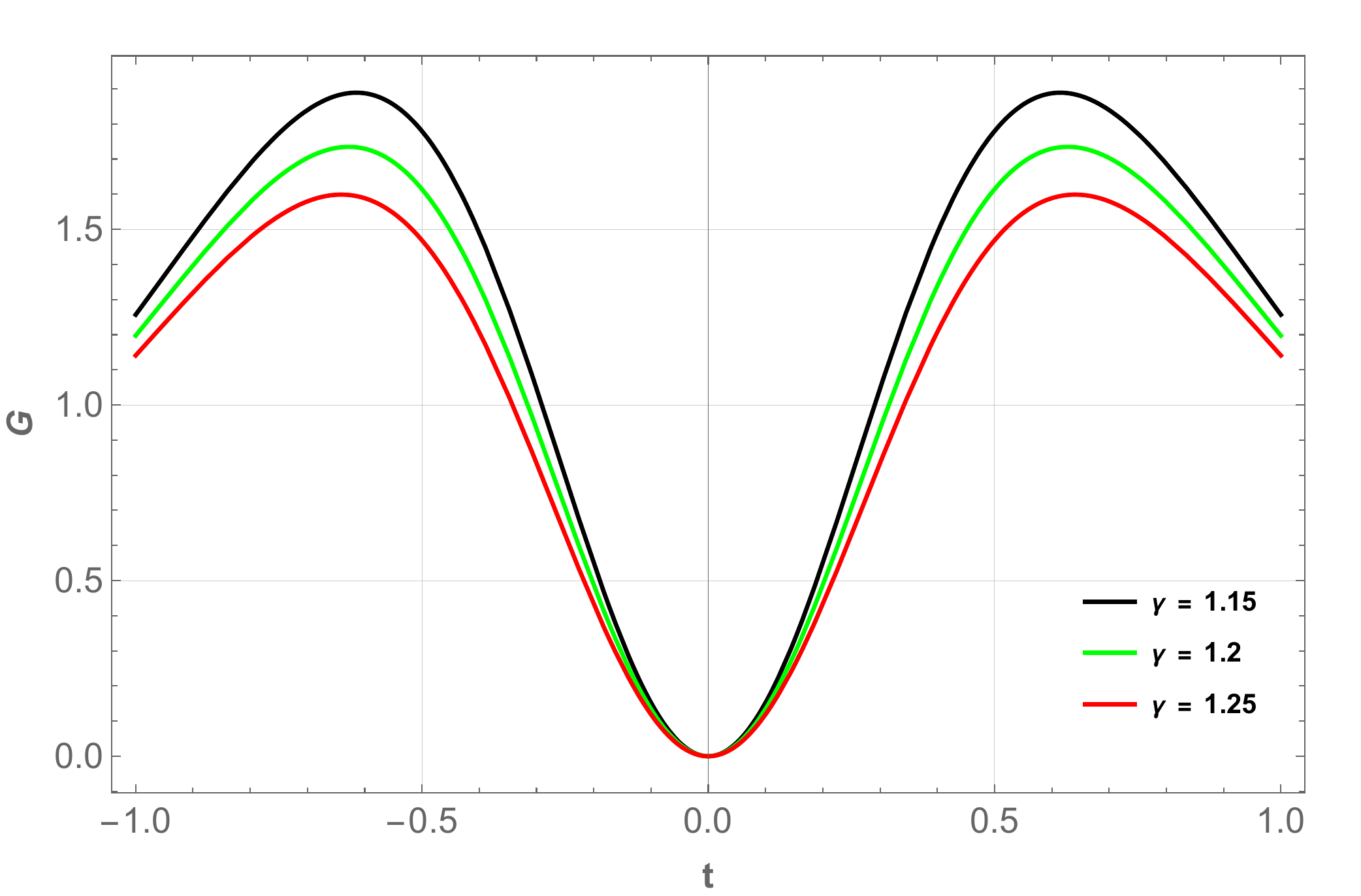}
\includegraphics[width=65mm]{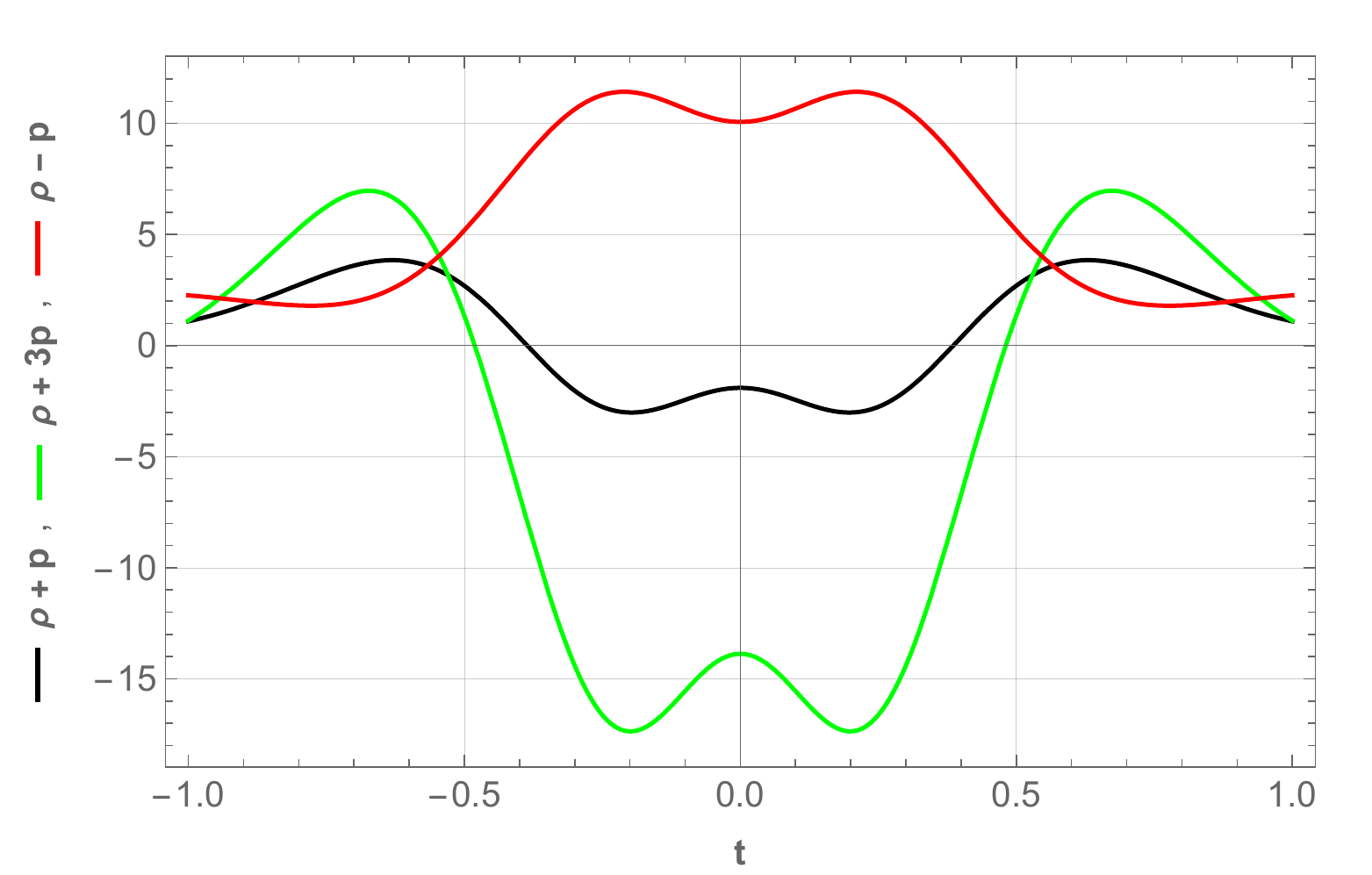}
\caption{Behavior of Gauss--Bonnet invariant (\textbf{left panel}) and energy conditions (\textbf{right panel}) versus cosmic time $t$, $\alpha=-0.30$, $\beta=0.15$ and $\lambda=1.01$.}
\label{FIG7}
\end{figure}

The graphical behavior of the Gauss--Bonnet invariant and the energy conditions are shown in Figure \ref{FIG7}. The bouncing behavior observed for the invariant, the violation of $\rho+3p$ and $\rho+p$, and satisfaction of $\rho-p$ was also obtained. The behavior of the Gauss--Bonnet term is visible, and the energy conditions are symmetric with regard to the bouncing epoch $t=0$. In the negative cosmic time domain, it decreased and continued to be negative, while in the positive cosmic time domain, it increased from negative values. It expands, even more, showing that, at $t \in (-0.648,0.648)$, it violates the $\rho+p$. Furthermore, given the $\rho+3p$, it lowers and decreases in the negative cosmic time domain at the bounce epoch. In contrast, it constantly increases in the positive cosmic time domain. The null energy condition is violated at the bounce epoch, and the apparent phenomenon is that the strong energy condition is also violated at the bounce epoch; however, it has been observed that it is violated throughout the evolution of the universe.
The ($j,s$) diagnostic pair can be calculated as
\begin{eqnarray}
j&=&\frac{(2 \lambda-1) \left(t^2 (\lambda-1)-3 \gamma \right)}{t^2}, \hspace{0.8cm} \nonumber \\
s&=&-\frac{(2 \lambda-1) \left[3 \gamma^2+6 \gamma t^2 (1-3 \lambda)+t^4 (\lambda-1) (3 \lambda-1)\right]}{t^4} \label{23}
\end{eqnarray}

\begin{figure} [H]
\centering
\includegraphics[width=65mm]{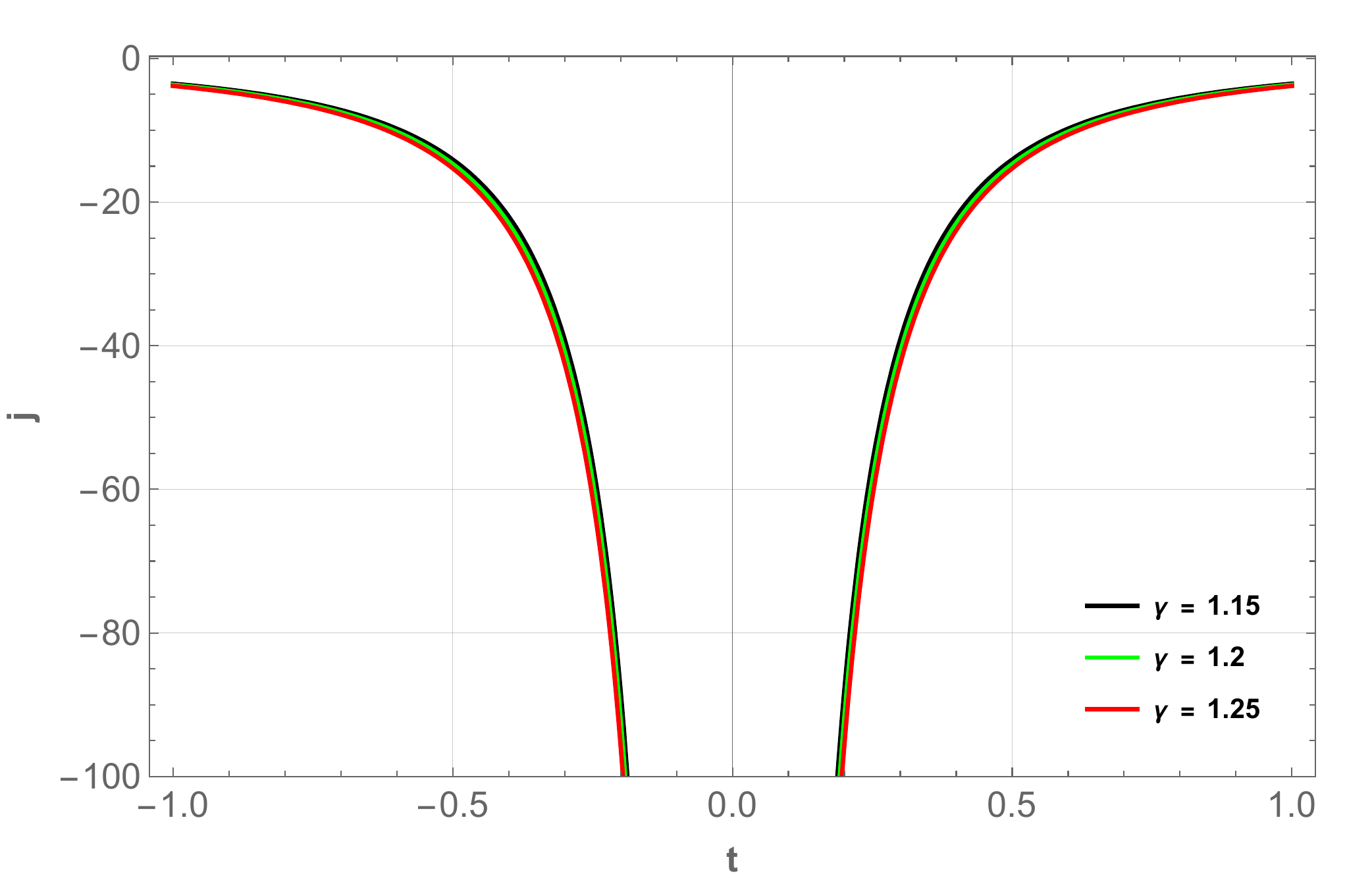}
\includegraphics[width=65mm]{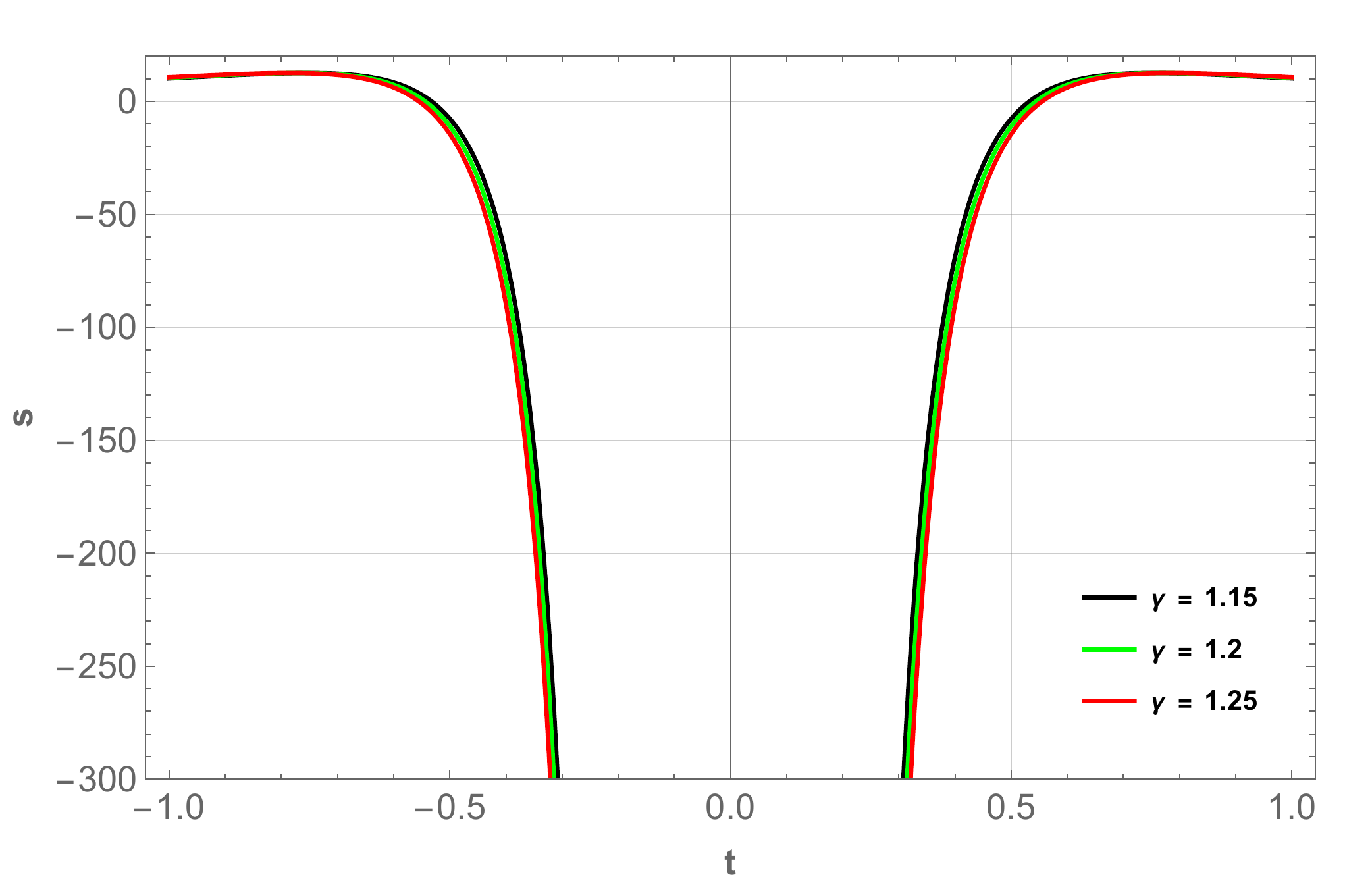}\\
\includegraphics[width=65mm]{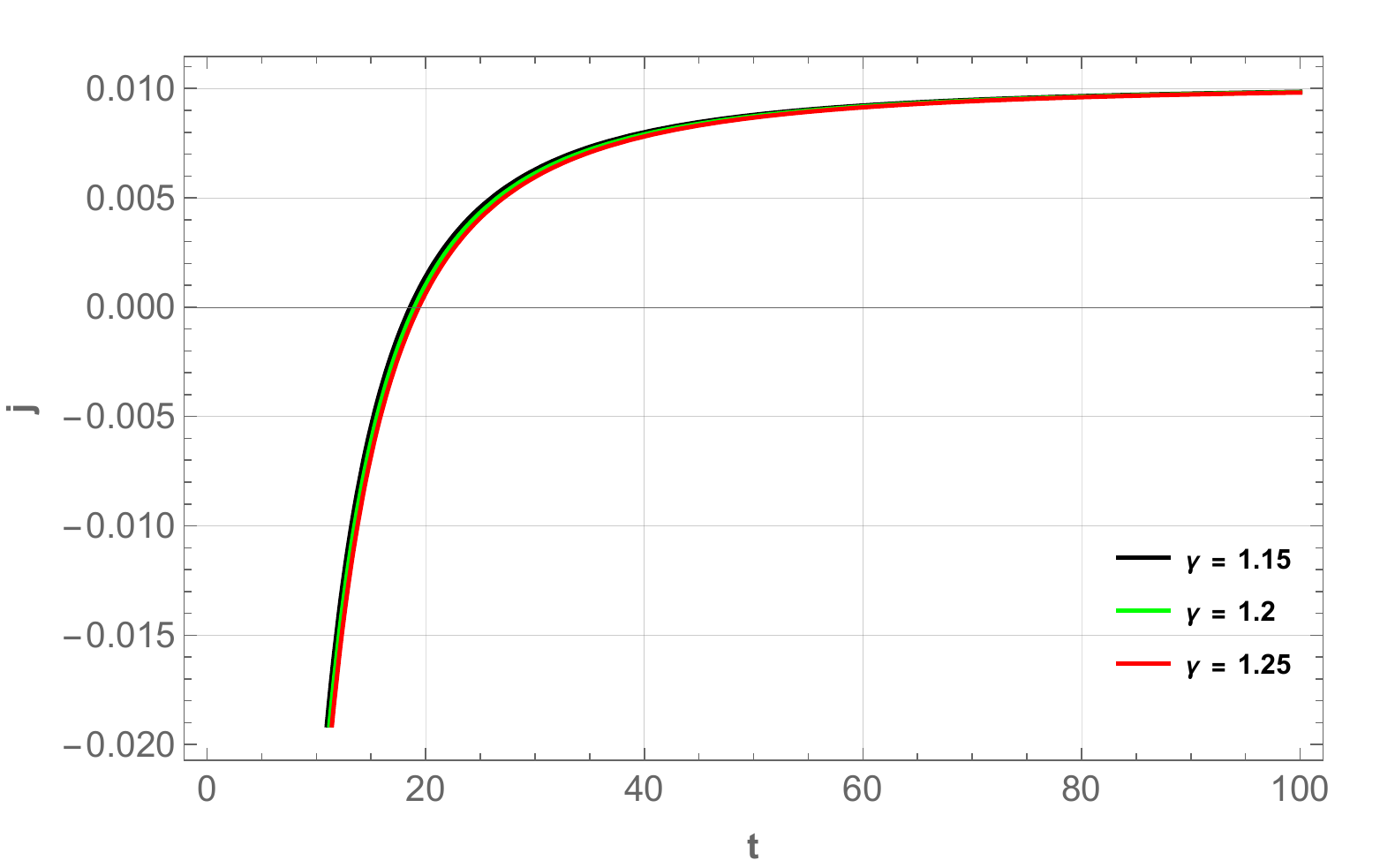}
\includegraphics[width=65mm]{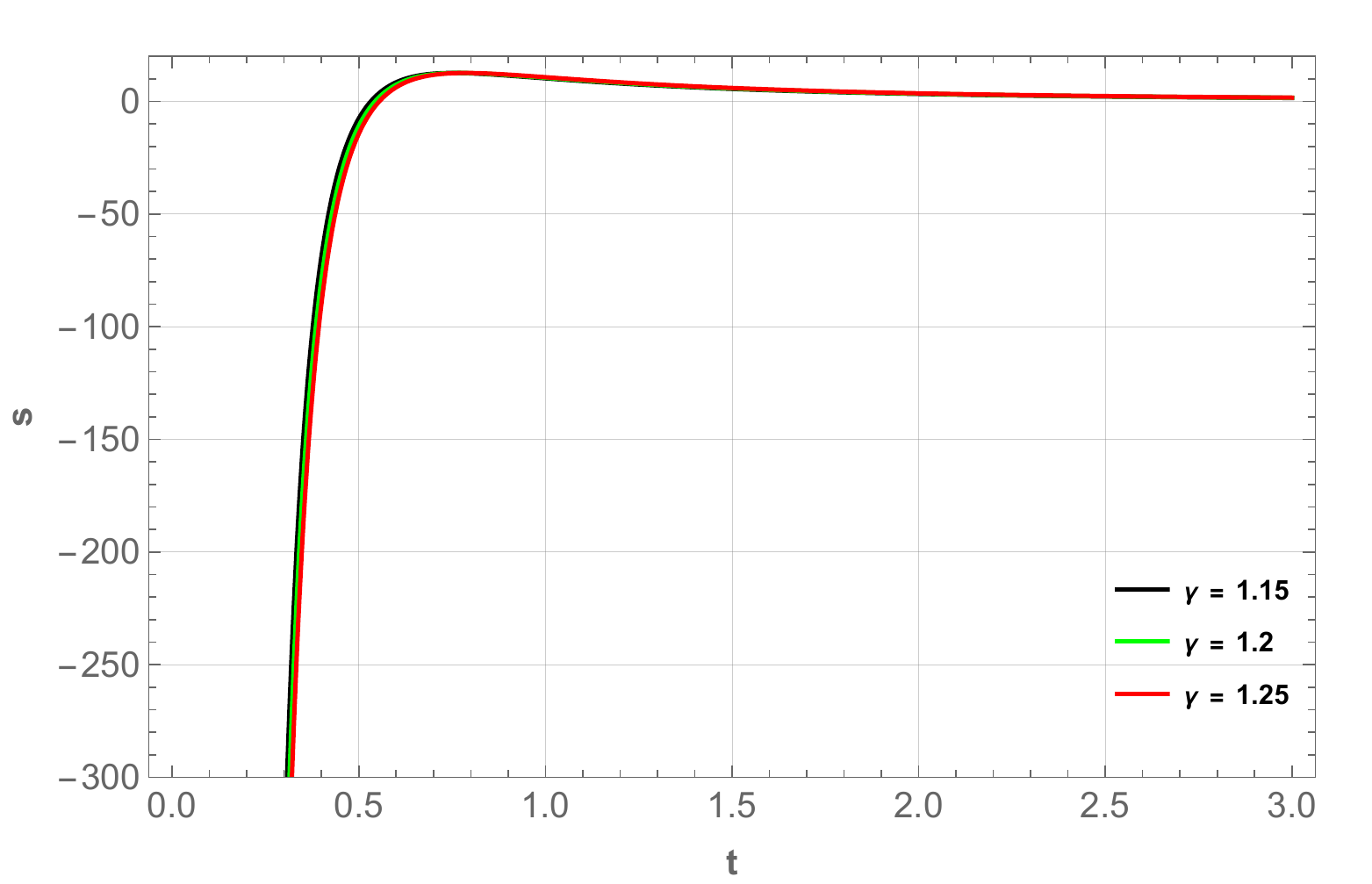}
\caption{Symmetric behavior of the jerk parameter (\textbf{above left panel}) and snap parameter (\textbf{above right panel}) versus cosmic time $t$. The jerk (\textbf{below left panel}) and snap (\textbf{below right panel}) parameter in positive time domain with $\lambda=1.01$.}
\label{FIG8}
\end{figure}

In Figure \ref{FIG8}, we plotted that the jerk and snap parameters are symmetric about the bouncing time $t=0$, which attains negative values. There is a unique bounce at the bouncing epoch for the jerk parameter, which is the third derivative of the scale factor.

\section{Scalar Perturbations} \label{SEC V}
Under linear homogeneous and isotropic perturbations, we shall investigate the stability of the bouncing cosmological models obtained in $F(R, \mathcal{G})$ gravity \cite{Dombriz12}. We shall use the pressureless dust FLRW background with a general explanation of $H(t) = H_{0}(t)$. The matter fluid is in the form of a perfect fluid with a constant EoS such that $p_{m} = \omega \rho_{m}$ and the matter-energy density $\rho_{m}$ obeys the standard continuity equation:
\begin{equation} \label{20}
\dot{\rho}_{m}+3H (1+\omega) \rho_{m}=0,
\end{equation}

Solving the continuity Equation \eqref{20}, the evolution of the matter-energy density can be described in terms of this specific solution
\begin{equation}\label{21}
\rho_{m{0}}(t)=\rho_{0} e^{-3 (1 + \omega_{m}) \int H_{0}(t)dt}
\end{equation}

The isotropic deviation of the baseline Hubble parameter and the matter over density is represented by $\delta(t)$ and $\delta_{m}(t)$, respectively. Now, we define the perturbation for the Hubble parameter and energy density as follows
\begin{equation}\label{22}
H(t)=H_{0}(t)\left(1+\delta (t)\right) \hspace{2cm} \rho_{m}(t)=\rho_{m{0}} \left(1+\delta_{m}(t)\right),
\end{equation}

We consider the Hubble parameter and the energy density around the arbitrary solutions $H_{0}(t)$ as perturbations \cite{Dombriz12}. We shall perform the perturbation analysis on the solution $H(t)=H_{0}(t)$, so that the function $F(R,\mathcal{G})$ may be represented in the powers of $R$ and $\mathcal{G}$ as

\begin{equation}\label{23}
F(R,\mathcal{G})=F^{0}+F_{R}^0(R-R_{0})+F_{\mathcal{G}}^0(\mathcal{G}-\mathcal{G}_0)+\mathcal{O}^2,
\end{equation}
where the subscript $0$ means the values of $F(R,\mathcal{G})$ and its derivatives $F_R$ and $F_\mathcal{G}$ are evaluated at $R=R_{0}$ and $\mathcal{G} = \mathcal{G}_{0}$. Although only the linear terms of the induced perturbations are examined, the $\mathcal{O}^2$ term contains all terms proportional to $R$ and square of $\mathcal{G}$ or higher powers that will be included in the equation and are ignored. Thus, by substituting Equations \eqref{22} and \eqref{23} in the FLRW background Equation \eqref{4} and the continuity Equation~\eqref{20}, we obtain the perturbation equations in terms of $\delta(t)$ and $\delta_{m}(t)$ in the form of the following differential equations
\begin{equation}\label{24}
c_{2}\ddot{\delta}(t) + c_{1}\dot{\delta}(t) + c_{0}\delta (t)=c_{m} \delta_{m}(t),
\end{equation}

The coefficients $c_{0}, c_{1}, c_{2}$, and $c_{m}$ (see Appendix \ref{sec:appendixA}) are explicitly dependent on the background of the $F(R, \mathcal{G})$ solution and its derivatives. In addition, once the matter continuity Equation~\eqref{20} is disturbed by expressions, a second perturbed equation is formed from Equation~\eqref{22}.~Thus,
\begin{equation}\label{25}
\dot{\delta}_{m}(t)+3H_0(t) \delta (t)=0.
\end{equation}
\begin{eqnarray} 
\delta (t)=\frac{-1}{2} \delta_m (t) \propto a(t)^{\frac{3}{2}}
\end{eqnarray}

We framed the model based on the functional $F(R,\mathcal{G}) = R + \alpha R^2+ \beta \mathcal{G}^2$. If we assume that GR will be retrieved from the current model at some point, we may have to ignore the contributions from the higher derivatives of the functional $F(R, \mathcal{G})$. Thus, using the perturbation approach in the equivalent FLRW equation, we obtain
\begin{eqnarray} \label{26}
-18 H_0(t)^2 \left(16 F_{\mathcal{G}\mathcal{G}}^0 H_0(t)^4 +F_{RR}^0 \right) \ddot{\delta}(t)-18 H_0(t) \big(-48 F_{\mathcal{G}\mathcal{G}}^0 H_0(t)^6-80 F_{\mathcal{G}\mathcal{G}}^0 H_0(t)^4 \dot{H}_0(t)\nonumber\\ -3 F_{RR}^0 H_0(t)^2-F_{RR}^0 \dot{H}_0(t)\big) \dot{\delta}(t) +6 \big[12 H_0(t)^4 (F_{RR}^0-36 F_{\mathcal{G}\mathcal{G}}^0 \dot{H}_0(t)^2)+192 F_{\mathcal{G}\mathcal{G}}^0 H_0(t)^8 \nonumber\\-1008 F_{\mathcal{G}\mathcal{G}}^0 H_0(t)^6 \dot{H}_0(t)-288 F_{\mathcal{G}\mathcal{G}}^0 H_0(t)^5 \ddot{H}_0 -H_0(t)^2 (F_R^0+21 F_{RR}^0 \dot{H}_0)\nonumber\\ -6 F_{RR}^0 H_0(t) \ddot{H}_0(t)+3 F_{RR}^0 \dot{H}_0(t)^2\big] \delta (t)=\kappa^2 \rho_{m0} \delta_m (t),
\end{eqnarray}

This is an algebraic relationship between geometrical and matter perturbations. As a result, matter perturbations ultimately dictate the whole perturbation surrounding a cosmological solution in GR. Now, we shall carry out the stability analysis for the two models discussed as presented in Section \ref{SEC V}.

\begin{figure} [H]
\centering
\includegraphics[width=68mm]{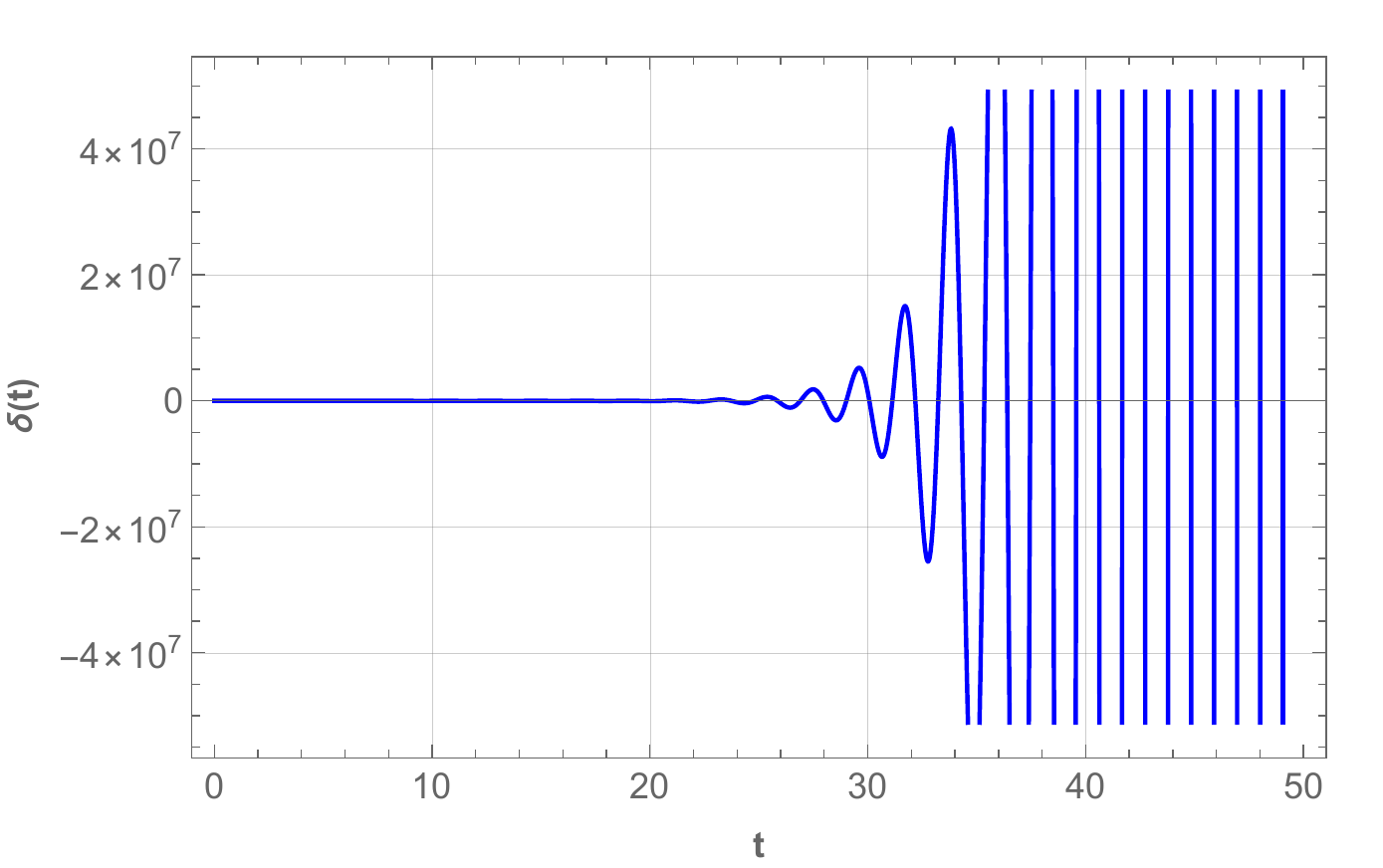}
\includegraphics[width=68mm]{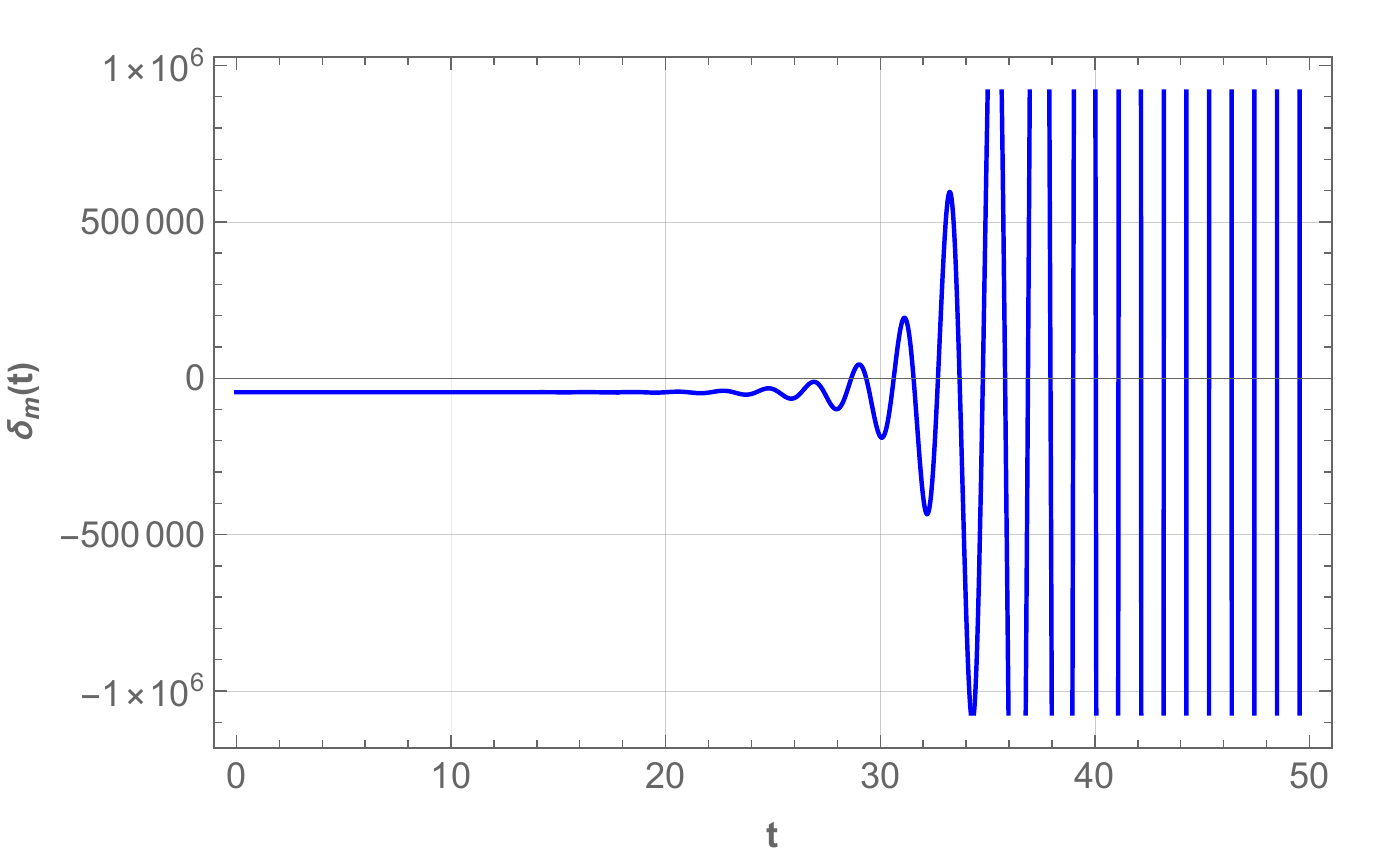}
\caption{$\delta(t)$ (\textbf{left panel}) and $\delta_{m}(t)$ (\textbf{right panel}) versus cosmic time for Model I. The parameter scheme: $\alpha=-0.30,\,\, \beta=0.15,\,\, \chi=0.15,\,\, \rho_0=1.2.$}
\label{FIG9}
\end{figure}

\begin{figure} [H]
\centering
\includegraphics[width=68mm]{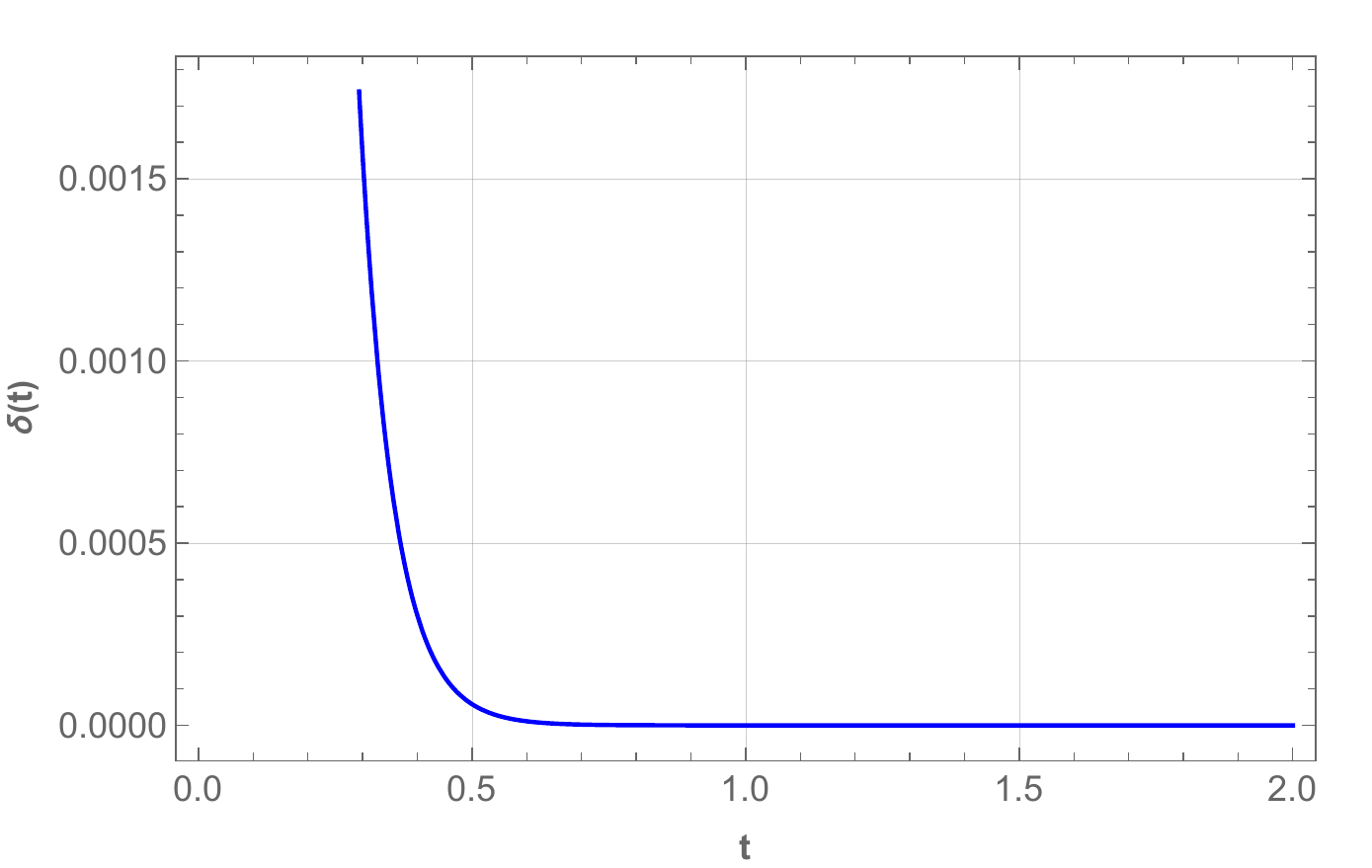}
\includegraphics[width=68mm]{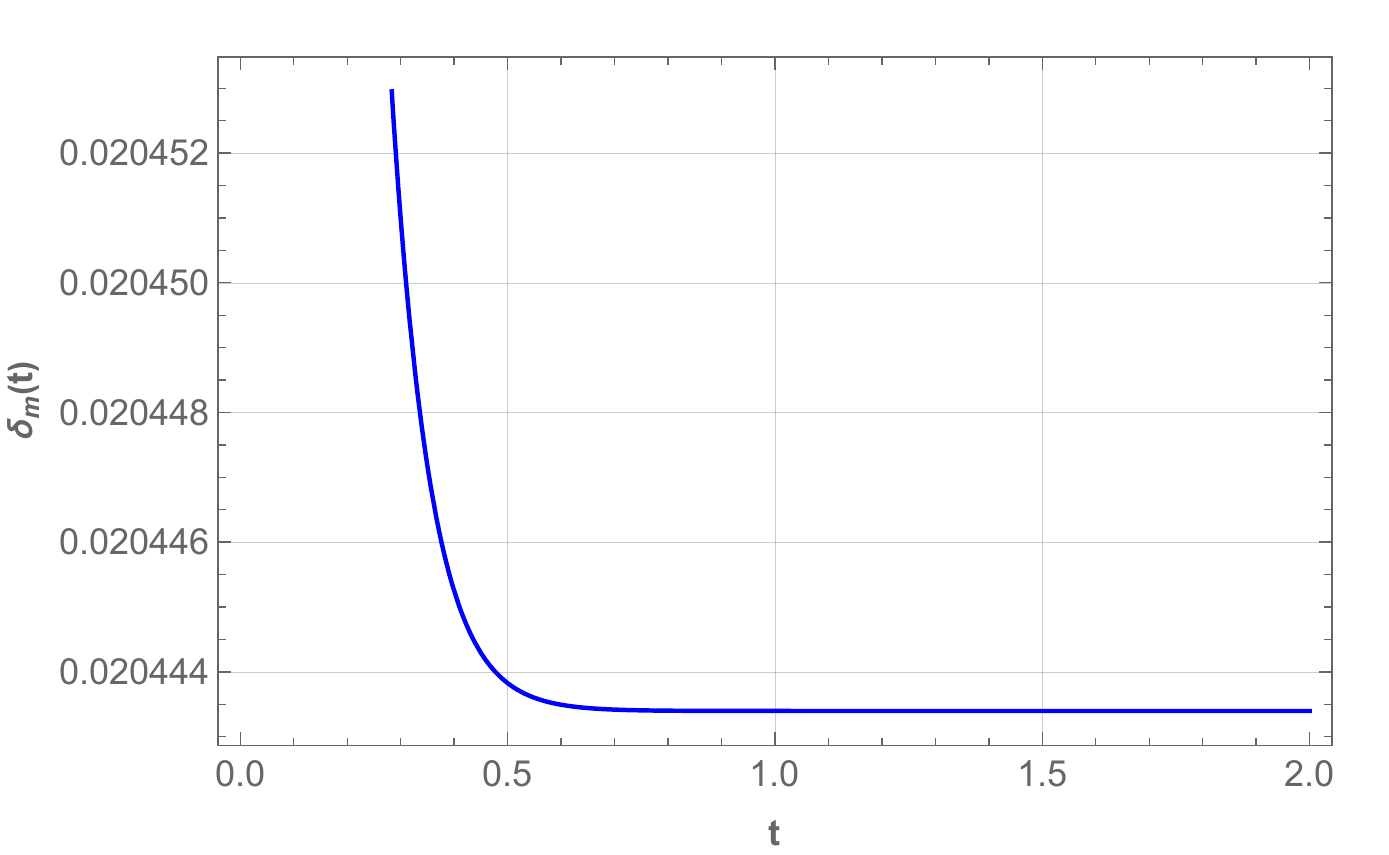}
\caption{$\delta(t)$ (\textbf{left panel}) and $\delta_{m}(t)$ (\textbf{right panel}) versus cosmic time for Model II. The parameter scheme: $\alpha = -0.30,\,\, \beta=0.15,\,\, \lambda=1.01,\,\, \rho_0=1.2.$}
\label{FIG10}
\end{figure}

The stability analysis of the model based on both bouncing scale factors is shown in Figures \ref{FIG9} and \ref{FIG10}. The visual assessment of the stability of the model is shown in both figures. For the chosen values of the parameters, the linear perturbations in the Hubble parameter and the energy density decrease as time passes; hence the stability of the model has been~ensured. 
The model \eqref{8} can be substituted in Equations \eqref{24} and \eqref{25} to obtain the essential perturbation equations for bouncing Model I. Figure \ref{FIG9} depicts the numerical scheme and evolution of $\delta(t)$ and $\delta_m (t)$. The figures indicate the oscillating behavior of $\delta(t)$ and $\delta_m (t)$, but the oscillations of $\delta(t)$ and $\delta_m (t)$ do not decay in the future. As a result, solutions are unstable since the complete perturbation surrounding a cosmological solution is defined entirely by matter perturbations. Figure \ref{FIG10} presents a graphical representation of the dynamical stability of the bouncing Model II. It demonstrates that $\delta(t)$ and $\delta_m (t)$ decay and yet indicate consistent behavior at later times. As a result, Model II is stable at later times. Throughout the evolution, the model remains stable for typical parameter values as $\delta(t)$ and $\delta_m (t)$ decrease and approach zero at a later time. We have investigated the linear homogeneous perturbation behaviors for the two bouncing models. The perturbation's progressive oscillating and decreasing behaviors exemplified the dynamical stability of the model with the increase in time, respectively, for the Model I and Model II.
\section{Conclusions} \label{SEC VI}
This paper investigated the cosmological behavior of a class of modified Gauss--Bonnet gravity models with two bouncing scale factors presented in the $F(R, \mathcal{G})$ theory of gravity. A specific functional form was chosen where the Ricci scalar and Gauss--Bonnet invariant are in quadratic form. Interestingly, the Gauss--Bonnet invariant for both the models shows the bouncing behavior. The violation of SEC and NEC, as prescribed for bouncing models, is achieved. The EoS parameter exhibits phantom-like behavior at the bounce point, then passes through the $\Lambda$CDM line before exhibiting quintessence-like behavior as we move away from the bounce point. The negative  $q$ in the models shows the accelerating behavior and the value of the EoS parameter to be well within the current cosmological observations. Finally, we examined the stability of the $F(R, \mathcal{G})$ model while accounting for the bouncing scale factor. For the Hubble parameter and energy density, we applied linear homogeneous perturbations. It is discovered that $F(R, \mathcal{G})$ gravity does not produce stable results for acceleration eras while failing to regenerate radiation and matter-dominated eras for both bouncing models. Figure \ref{FIG9} shows the oscillating behavior of $\delta(t)$ and $\delta_m(t)$; however, the oscillations of $\delta(t)$ and $\delta_m(t)$ do not decay in the future. As a result, the solutions are unstable since the whole perturbation around a cosmological solution is exclusively determined by matter perturbations. Figure \ref{FIG10} represents the dynamical stability of bouncing Model II. This shows that $\delta(t)$ and $\delta_m(t)$ decrease, but it also indicates consistent behavior at later times. As a result, Model II is stable at later times.
\appendix
\section[\appendixname~\thesection]{}
\begin{eqnarray*}
\frac{1}{6} c_0 &=&-18432 H_0^{10} \dot{H}_0 F_{3 \mathcal{G}}^0 +3 \dot{H}_0^2 F_{RR}^0+192 H_0^8 (F_{\mathcal{G}\mathcal{G}}^0-24 \dot{H}_0 (3 F_{R\mathcal{G}\mathcal{G}}^0+5 F_{3\mathcal{G}}^0 \dot{H}_0))\\ & &+48 H_0^6 \left(2 F_{R\mathcal{G}}^0-3 \dot{H} (7 F_{\mathcal{G}\mathcal{G}}^0+24 F_{RR\mathcal{G}}^0+88 F_{R\mathcal{G}\mathcal{G}}^0 \dot{H}_0+48 F_{3\mathcal{G}}^0 \dot{H}_0^2) \right)\\ & &+12 H_0^4 \left(F_{RR}^0-4 \dot{H}_0 (7 F_{RG}^0+6 F_{3R}^0+3 \dot{H}_0 (3 F_{\mathcal{G}\mathcal{G}}^0+14 F_{RR\mathcal{G}}^0+16 F_{R \mathcal{G}\mathcal{G}}^0 \dot{H}_0))  \right)\\ & &-H_0^2 \left[F_R^0+3\dot{H}_0 (7 F_{RR}^0+ 8 \dot{H}_0 (2 F_{R\mathcal{G}}^0 + 3 F_{3R}^0 +6 F_{RR\mathcal{G}}^0 \dot{H}_0)) \right]-4608 H_0^9 \ddot{H}_0 F_{3\mathcal{G}}^0\\ & &-3456 H_0^7 \ddot{H}_0 (F_{R\mathcal{G}\mathcal{G}}^0+ F_{3\mathcal{G}}^0 \dot{H}_0)-288 H_0^5 \ddot{H}_0 (F_{\mathcal{G}\mathcal{G}}^0+3 F_{RR\mathcal{G}}^0+7 F_{R\mathcal{G}\mathcal{G}}^0 \dot{H}_0)\\ & &-6 H_0 \ddot{H}_0 (F_{RR}^0+3 F_{3R}^0 \dot{H}_0)-24 H_0^3 \ddot{H}_0 (4 F_{R \mathcal{G}}^0+3 (F_{3R}^0+5 F_{RR\mathcal{G}}^0 \dot{H}_0)),\\
c_1 &=&-18 H_0 \big[1536 H_0^8 \dot{H}_0 F_{3\mathcal{G}}^0-F_{RR}^0 \dot{H}_0 -48 H_0^6 \left(F_{\mathcal{G}\mathcal{G}}^0+8 \dot{H}_0 (3 F_{R\mathcal{G}\mathcal{G}}^0+2 \dot{H}_0 F_{3\mathcal{G}}^0) \right)\\ & & -384 H_0^7 \ddot{H}_0 F_{3 \mathcal{G}}^0 -8 H_0^4 \left(3 F_{RG}^0+2 \dot{H}_0 (5 F_{\mathcal{G}\mathcal{G}}^0+18 F_{RR\mathcal{G}}^0+24 \dot{H}_0 F_{R\mathcal{G}\mathcal{G}}^0) \right)-3 H_0^2 \big(F_{RR}^0 \\ & & +8 \dot{H}_0 (F_{R \mathcal{G}}^0+ F_{3R}^0 +2 F_{RRG}^0 \dot{H}_0) \big)-288 H_0^5 \ddot{H}_0 F_{R\mathcal{G}\mathcal{G}}^0-72 H_0^3 \ddot{H}_0 F_{RR\mathcal{G}}^0-6 H_0 \ddot{H}_0 F_{3R}^0 \big],\\
c_2 &=&-18 H_0^2 (16 F_{\mathcal{G}\mathcal{G}}^0 H_0^4+8 F_{\mathcal{G} R}^0 H_0^2+F_{RR}^0),\\
c_m&=&\kappa^2 \rho_{m0}(t).
\end{eqnarray*}


\begin{thebibliography}{99}
\section*{References}
\bibitem{Riess98} Riess, A.G.; et al. Observational Evidence from Supernovae for an Accelerating Universe and a Cosmological Constant. \href{https://doi.org/10.1086/300499}{\textit{Astron. J.} \textbf{1998}, 116, 1009.} 

\bibitem{Perlmutter99} Perlmutter, S.; et al. Measurements of $\Omega$ and $\Lambda$ from 42 High-Redshift Supernovae. \href{https://doi.org/10.1086/307221}{\textit{Astrophys. J.} \textbf{1999}, 517, 565.}

\bibitem{Bennett03} Bennett, C.L.; et al. First-Year Wilkinson Microwave Anisotropy Probe (WMAP)* Observations: Foreground Emission. \href{https://doi.org/10.1086/377252}{\textit{Astrophys. J. Suppl. Ser.} \textbf{2003}, 148, 1.}

\bibitem{Spergel03} Spergel, S.; et al. First-Year Wilkinson Microwave Anisotropy Probe (WMAP)* Observations: Determination of Cosmological Parameters. \href{https://doi.org/10.1086/377226}{\textit{Astrophys. J.} \textbf{2003}, 148, 175.}

\bibitem{Spergel07} Spergel, S.; et al. Three-Year Wilkinson Microwave Anisotropy Probe (WMAP) Observations: Implications for Cosmology. \href{https://doi.org/10.1086/513700}{\textit{Astrophys. J.} \textbf{2007}, 170, 377.}

\bibitem{Percival10} Percival, W.J.; et al. Baryon acoustic oscillations in the Sloan Digital Sky Survey Data Release 7 galaxy sample. \href{https://doi.org/10.1111/j.1365-2966.2009.15812.x}{\textit{Mon. Not. R. Astron. Soc.} \textbf{2010}, 401, 2148.}

\bibitem{Ade14} Ade, P.A.R.; et al. Detection of B-Mode Polarization at Degree Angular Scales by BICEP2. \href{https://doi.org/10.1103/PhysRevLett.112.241101}{\textit{Phys. Rev. Lett.} \textbf{2014}, 112, 241101.}

\bibitem{Nojiri03a} Nojiri, S.; Odintsov, S.D. Modified gravity with negative and positive powers of curvature: Unification of inflation and cosmic acceleration. \href{https://doi.org/10.1103/PhysRevD.68.123512}{\textit{Phys. Rev. D} \textbf{2003}, 68, 123512.}

\bibitem{Carroll04} Carroll, S.M.; et al. Is cosmic speed-up due to new gravitational physics? \href{https://doi.org/10.1103/PhysRevD.70.043528} {\textit{Phys. Rev. D} \textbf{2004}, 70, 043528.}

\bibitem{Barragan09} Barragan, C.; Olmo, G.J.; Sanchis-Alepuz, H. Bouncing cosmologies in Palatini $f(R)$ gravity \href{http://dx.doi.org/10.1103/PhysRevD.80.024016}{\textit{Phys. Rev. D} \textbf{2019}, 80, 024016.}

\bibitem{Harko11} Harko, T.; et al. $f(R,T)$ gravity \href{http://dx.doi.org/10.1103/PhysRevD.84.024020}{\textit{Phys. Rev. D} \textbf{2011}, 84, 024020.}

\bibitem{Cai11} Cai, Y.F.; et al. Matter bounce cosmology with the $f(T)$ gravity \href{https://doi.org/10.1088/0264-9381/28/21/215011}{\textit{Class. Quant. Grav.} \textbf{2011}, 28, 215011.}

\bibitem{Abedi18} Abedi, H.; et al. Effective gravitational coupling in modified teleparallel theories \href{https://doi.org/10.1103/PhysRevD.97.084008}{\textit{Phys. Rev. D} \textbf{2018}, 97, 084008.}

\bibitem{Xu19} Xu, Y.; et al. $f(Q,T)$ gravity \href{https://doi.org/10.1140/epjc/s10052-019-7207-4}{\textit{Eur. Phys. J. C} \textbf{2019}, 79, 708.}

\bibitem{Nojiri05} Nojiri, S.; Odintsov, S.D. Modified Gauss-Bonnet theory as gravitational alternative for dark energy \href{https://doi.org/10.1016/j.physletb.2005.10.010}{\textit{Phys. Lett. B} \textbf{2005}, 631, 6.}

\bibitem{Cai07} Cai, Y.F.;  et al. Bouncing universe with Quintom matter \href{https://doi.org/10.1088/1126-6708/2007/10/071}{\textit{JHEP} \textbf{2007}, 10, 071.} 

\bibitem{Peter02} Peter, P.; Pinto-Neto, N. Primordial perturbations in a nonsingular bouncing universe model \href{https://doi.org/10.1103/PhysRevD.66.063509}{\textit{Phys. Rev. D} \textbf{2002}, 66, 063509.}

\bibitem{Lin11} Lin, C.; Brandenberger, R. H.; Perreault, L.L. A matter bounce by means of ghost condensation \href{https://doi.org/10.1088/1475-7516/2011/04/019}{\textit{J. Cosmol. Astropart. Phys.} \textbf{2011}, 04, 019.} 

\bibitem{Fabris03} Fabris, J.C.; et al. Regular cosmological bouncing solutions in low energy effective action from string theories \href{https://doi.org/10.1103/PhysRevD.67.124003}{\textit{Phys. Rev. D} \textbf{2003}, 67, 124003.}

\bibitem{Qiu11} Qiu, T.; et al. Bouncing Galileon cosmologies \href{https://doi.org/10.1088/1475-7516/2011/10/036}{\textit{J. Cosmol. Astropart. Phys.} \textbf{2011}, 10, 036.}

\bibitem{Kounnas12} Kounnas, C.; et al. Thermal duality and non-singular cosmology in d-dimensional superstrings \href{https://doi.org/10.1016/j.nuclphysb.2011.10.010}{\textit{Nucl. Phys. B} \textbf{2012}, 855, 280.}

\bibitem{Novello08} Novello, M.; Perez Bergliaffa, S.E. Bouncing cosmologies \href{https://doi.org/10.1016/j.physrep.2008.04.006}{\textit{Phys. Rep.} \textbf{2008}, 463, 127.}

\bibitem{Elizalde15} Elizalde, E.; Haro, J.; Odintsov, S.D.; Quasimatter domination parameters in bouncing cosmologies \href{https://doi.org/10.1103/PhysRevD.91.063522}{\textit{Phys. Rev. D} \textbf{2015}, 90, 063522.}

\bibitem{Peebles70} Peebles, P.J.E.; Yu, J.T. Primeval Adiabatic Perturbation in an Expanding Universe \href{https://ui.adsabs.harvard.edu/link_gateway/1970ApJ...162..815P/doi:10.1086/150713}{\textit{Astrophys. J.} \textbf{1970}, 162, 815.}

\bibitem{Sunyaev70} Sunyaev, R.A.; Zeldovich, Y.B. Small-scale fluctuations of relic radiation. \href{https://doi.org/10.1007/BF00653471}{\textit{Astrophys. Space Sci.} \textbf{1970}, 7, 3.}

\bibitem{Battefeld15} Battefeld, D.; Peter, P. A critical review of classical bouncing cosmologies. \href{https://doi.org/10.1016/j.physrep.2014.12.004}{\textit{Phys. Rep.} \textbf{2015}, 571, 1.}

\bibitem{Barrau17} Barrau, A.; et al. Bouncing black holes in quantum gravity and the Fermi gamma-ray excess \href{https://doi.org/10.1016/j.physletb.2017.05.040}{\textit{Phys. Lett. B} \textbf{2017}, 772, 58.}

\bibitem{Bojowald01} Bojowald, M. Absence of a Singularity in Loop Quantum Cosmology \href{https://doi.org/10.1103/PhysRevLett.86.5227}{\textit{Phys. Rev. Lett.} \textbf{2011}, 86, 5227.}

\bibitem{Odintsov14} Odintsov, S.D.; Oikonomou, V.K. Matter Bounce Loop Quantum Cosmology from $F(R)$ Gravity \href{https://arxiv.org/abs/1410.8183v2}{\textit{Phys. Rev. D} \textbf{2014}, 90, 124083.}

\bibitem{Khoury01} Khoury, J.; et al. Ekpyrotic universe: Colliding branes and the origin of the hot big bang \href{https://doi.org/10.1103/PhysRevD.64.123522}{\textit{Phys. Rev. D} \textbf{2001}, 64, 123522.}

\bibitem{Gasperini03} Gasperini, M.; Giovannini, M.; Veneziano, G. Perturbations in a non-singular bouncing Universe \href{https://doi.org/10.1016/j.physletb.2003.07.028}{\textit{Phys. Lett. B} \textbf{2003}, 569, 113.}

\bibitem{Nojiri03} Nojiri, S.; Saridakis, E.N. Phantom without ghost \href{https://doi.org/10.1007/s10509-013-1509-z}{\textit{Astrophys. Space Sci.} \textbf{2003}, 347, 221.}

\bibitem{Saridakis09} Saridakis, E.N. Cyclic universes from general collisionless braneworld models \href{https://doi.org/10.1016/j.nuclphysb.2008.09.022}{\textit{Nucl. Phys. B} \textbf{2009}, 808, 224.}

\bibitem{Brandenberger09}  Brandenberger, R. Matter bounce in Horava-Lifshitz cosmology \href{http://dx.doi.org/10.1103/PhysRevD.80.043516}{\textit{Phys. Rev. D} \textbf{2009}, 80, 043516.}

\bibitem{Cai09} Cai, Y.F.; Saridakis, E.N. Non-singular cosmology in a model of non-relativistic gravity \href{https://doi.org/10.1088/1475-7516/2009/10/020}{\textit{J. Cosmol. Astropart. Phys.} \textbf{2009}, 10, 020.}

\bibitem{Agrawal21} Agrawal, A.S.; et al. Matter bounce scenario and the dynamical aspects in $f(Q,T)$ gravity \href{https://doi.org/10.1016/j.dark.2021.100863}{\textit{Phys. Dark Universe} \textbf{2021}, 33, 100863.}

\bibitem{Cai16} Cai, Y.F.; et al. Searching for a matter bounce cosmology with low redshift observations \href{https://doi.org/10.1103/PhysRevD.93.043546}{\textit{Phys. Rev. D} \textbf{2016}, 93, 043546.}

\bibitem{Shabani18} Shabani, H.; Ziaie, A.H. Bouncing cosmological solutions from $f(R,T)$ gravity \href{https://doi.org/10.1140/epjc/s10052-018-5886-x}{\textit{Eur. Phys. J. C} \textbf{2018}, 78, 397.}

\bibitem{Mishra19} Mishra, B.; Ribeiro, G.; Moraes, P.H.R.S. De Sitter and bounce solutions from anisotropy in extended gravity cosmology \href{https://doi.org/10.1142/S0217732319503218}{\textit{Mod. Phys. Lett. A} \textbf{2019}, 34, 1950321.}

\bibitem{Tripathy21} Tripathy, S.K.; et al. Bouncing universe models in an extended gravity theory \href{https://doi.org/10.1016/j.cjph.2021.03.026}{\textit{Chin. J. Phys.} \textbf{2021}, 71, 610.}

\bibitem{Agrawal22p} Agrawal, A.S.; et al. Role of extended gravity theory in matter bounce dynamics \href{https://doi.org/10.1088/1402-4896/ac49b2}{\textit{Phys. Scr.} \textbf{2022}, 97, 025002.}

\bibitem{Agrawal22f} Agrawal, A.S.; et al. Bouncing Cosmology in Extended Gravity and Its Reconstruction as Dark Energy Mode \href{https://doi.org/10.1002/prop.202100065}{\textit{Fortschr. Phys.} \textbf{2022}, 70, 2100065.}

\bibitem{Amani16} Amani, A.R. The bouncing cosmology with $F(R)$ gravity and its reconstructing \href{https://doi.org/10.1142/S0218271816500711}{\textit{Int. J. Mod. Phys. D} \textbf{2016}, 25, 1650071.}

\bibitem{Nojiri16} Nojiri, S.; Odintsov, S.D.; Oikonomou, V.K. Bounce universe history from unimodular $F(R)$ gravity \href{https://doi.org/10.1103/PhysRevD.93.084050}{\textit{Phys. Rev. D} \textbf{2016}, 93, 084050.}

\bibitem{Ilyas21} Ilyas, M.; Rahman, W.U. Bounce cosmology in $f(R)$ gravity \href{https://doi.org/10.1140/epjc/s10052-021-08955-7}{\textit{Eur. Phys. J. C} \textbf{2021}, 81, 160.}

\bibitem{Amoros13} Amoros, J.; de Haro, J.; Odintsov, S.D. Bouncing loop quantum cosmology from $F(T)$ gravity \href{http://dx.doi.org/10.1103/PhysRevD.87.104037}{\textit{Phys. Rev. D} \textbf{2013}, 87, 104037.}

\bibitem{Caruana20} Caruana, M.; Farrugia, G.; Levi Said, J. Cosmological bouncing solutions in $f(T, B)$ gravity \href{https://doi.org/10.1140/epjc/s10052-020-8204-3}{\textit{Eur. Phys. J. C} \textbf{2020}, 80, 640.} 

\bibitem{Odintsov20} Odintsov, S.D; Oikonomou, V.K; Paul, T. Bottom-up reconstruction of non-singular bounce in $F(R)$ gravity from observational indices \href{https://doi.org/10.1016/j.nuclphysb.2020.115159}{\textit{Nucl. Phys. B} \textbf{2020}, 959, 115159.}

\bibitem{Karimzadeh19} Karimzadeh, S.; Shojaee, R. Phantom-Like Behavior in Modified Teleparallel Gravity \href{https://doi.org/10.1155/2019/4026856}{\textit{Adv. High Energy Phys.} \textbf{2019}, 8, 4026856.}

\bibitem{Duchaniya22} Duchaniya, L.K.; Lohakare, S.V.; Mishra, B.; Tripathy, S.K. Dynamical stability analysis of accelerating $f(T)$ gravity models \href{https://doi.org/10.1140/epjc/s10052-022-10406-w}{\textit{Eur. Phys. J. C} \textbf{2022}, 82, 448.}

\bibitem{Barrow88} Barrow, J.D.; Cotsakis, S. Inflation and the conformal structure of higher-order gravity theories \href{https://doi.org/10.1016/0370-2693(88)90110-4}{\textit{Phys. Lett. B} \textbf{1988}, 214, 515.}

\bibitem{Capozziello04} Capozziello, S.; et al. Can higher order curvature theories explain rotation curves of galaxies? \href{https://doi.org/10.1016/j.physleta.2004.04.081}{\textit{Phys. Lett. A} \textbf{2004}, 326, 292.}

\bibitem{Elizalde10} Elizalde, E.; et al. $\Lambda$CDM epoch reconstruction from $F(R,\mathcal{G})$ and modified Gauss-Bonnet gravities \href{https://doi.org/10.1088/0264-9381/27/9/095007}{\textit{Class. Quantum Grav.} \textbf{2010}, 27, 095007.}

\bibitem{Dombriz12} de la Cruz-Dombriz, A.; Saez-Gomez, D. On the stability of the cosmological solutions in $f(R, G)$ gravity \href{https://doi.org/10.1088/0264-9381/29/24/245014}{\textit{Class. Quantum Grav.} \textbf{2012}, 29, 245014.}

\bibitem{Cognola06} Cognola, G.; et al. Dark energy in modified Gauss-Bonnet gravity: Late-time acceleration and the hierarchy problem \href{http://dx.doi.org/10.1103/PhysRevD.73.084007}{\textit{Phys. Rev. D} \textbf{2006}, 73, {084007}.}

\bibitem{Felice09} De Felice, A.; Suyama, T. Vacuum structure for scalar cosmological perturbations in modified gravity models \href{https://doi.org/10.1088/1475-7516/2009/06/034}{\textit{J. Cosmol. Astropart. Phys.} \textbf{2009}, 0906, 034.}

\bibitem{Felice10} De Felice, A.; Gerard, J.M.; Suyama, T. Cosmological perturbation in $f(R,G)$ theories with a perfect fluid \href{http://dx.doi.org/10.1103/PhysRevD.82.063526}{\textit{Phys. Rev. D} \textbf{2010}, 82, 063526.}

\bibitem{Felice11} De Felice, A.; Suyama, T.; Tanaka, T. Stability of Schwarzschild-like solutions in $f(R,G)$ gravity models \href{http://dx.doi.org/10.1103/PhysRevD.83.104035}{\textit{Phys. Rev. D} \textbf{2011}, 83, 104035.}

\bibitem{Makarenko13} Makarenko, A.N.; Obukhov, V.V.; Kirnos, I. V. From Big to Little Rip in modified $F(R,G)$ gravity \href{https://doi.org/10.1007/s10509-012-1240-1}{\textit{Astrophys. Space Sci.} \textbf{2013}, 343, 481.}

\bibitem{Laurentis15} De Laurentis, M.; Paolella, M.; Capozziello, S. Cosmological inflation in $F(R,G)$ gravity \href{https://doi.org/10.1103/PhysRevD.91.083531}{\textit{Phys. Rev. D} \textbf{2015}, 91, {083531}.}

\bibitem{Martino20} de Martino, I.; De Laurentis, M.; Capozziello, S. Tracing the cosmic history by Gauss-Bonnet gravity \href{https://doi.org/10.1103/PhysRevD.102.063508}{\textit{Phys. Rev. D} \textbf{2020}, 102, {063508}.}

\bibitem{Haro12} Haro, J. Future singularity avoidance in phantom dark energy models \href{http://doi.org/10.1088/1475-7516/2012/07/007}{\textit{J. Cosmol. Astropart. Phys.} \textbf{2012}, 07, 007.}

\bibitem{Haro14} Haro, J.; Amor\'os, J. Viability of the matter bounce scenario in $F(T)$ gravity and Loop Quantum Cosmology for general potentials \href{http://doi.org/10.1088/1475-7516/2014/12/031} {\textit{J. Cosmol. Astropart. Phys.} \textbf{2014},  12, 031.}

\bibitem{Cai10} Cai, Y.F.; et al. Quintom cosmology: Theoretical implications and observations \href{https://doi.org/10.1016/j.physrep.2010.04.001}{\textit{Phys. Rep.} \textbf{2010}, 493, 1.}

\bibitem{Caldwell02} Caldwell, R. A phantom menace? Cosmological consequences of a dark energy component with super-negative equation of state \href{https://doi.org/10.1016/S0370-2693(02)02589-3}{\textit{Phys. Lett. B} \textbf{2002}, 545, 23.}

\bibitem{Steinhardt99} Steinhardt, P.; Wang, L.; Zlatev, I. Cosmological tracking solutions \href{https://doi.org/10.1103/PhysRevD.59.123504}{\textit{Phys. Rev. D} \textbf{1999}, 59, 123504.}

\bibitem{Wang06} Wang, B.; Gong, Y.; Abdalla, E. Thermodynamics of an accelerated expanding universe \href{http://dx.doi.org/10.1103/PhysRevD.74.083520}{\textit{Phys. Rev. D} \textbf{2006}, 74, 083520.}

\bibitem{Nojiri11} Nojiri, S.; Odintsov, S.D. Unified cosmic history in modified gravity: From $F(R)$ theory to Lorentz non-invariant models \href{https://doi.org/10.1016/j.physrep.2011.04.001}{\textit{Phys. Rep.}  \textbf{2011}, 505, 59.}

\bibitem{Capozziello11} Capozziello, S.; De Laurentis, M. Extended Theories of Gravity \href{https://doi.org/10.1016/j.physrep.2011.09.003}{\textit{Phys. Rep.} \textbf{2011}, 509, 167.}

\bibitem{Sahni03} Sahni, V.; et. al. Statefinder-A new geometrical diagnostic of dark energy \href{https://doi.org/10.1134/1.1574831}{\textit{JETP Lett.} \textbf{2003}, 77, 201.}

\bibitem{Alam03} Alam, U.; et. al. Exploring the expanding Universe and dark energy using the statefinder diagnostic \href{https://doi.org/10.1046/j.1365-8711.2003.06871.x}{\textit{Mon. Not. R. Astron. Soc.} \textbf{2003}, 344, 1057.} 

\bibitem{Abdussattar11} Abdussattar; Prajapati, S.R. Role of deceleration parameter and interacting dark energy in singularity avoidance \href{https://doi.org/10.1007/s10509-010-0461-4}{\textit{Astrophys. Space Sci.} \textbf{2011}, 331, 657.}

\end{thebibliography}
\end{document}